\documentclass[reprint,superscriptaddress,showkeys,showpacs]{revtex4-2}
\usepackage{verbatim}
\usepackage{color}
\usepackage{subfigure}
\usepackage[]{graphicx}
\usepackage{dcolumn}
\usepackage{bm}
\usepackage{blindtext}
\usepackage{floatrow}
\usepackage{makeidx}
\usepackage{epsfig}
\usepackage{hyperref}
\usepackage{array}
\usepackage{epstopdf}
\usepackage{braket}
\usepackage{multirow}
\usepackage{mathtools}
\usepackage{amsmath,amssymb,amsfonts}
\usepackage{fancyhdr}
\usepackage{multirow}
\emergencystretch=\maxdimen
\newcommand{\RNum}[1]
{\uppercase\expandafter{\romannumeral #1\relax}}
\newcommand{\Rnum}[1]
{\lowercase\expandafter{\romannumeral #1\relax}}
\def\qe{\textsc{Quantum ESPRESSO}}

\makeindex
\begin{document}
	\preprint{APS/123-QED}
	\title{Structural, electronic, and optical properties of 2D $\alpha$-graphdiyne from first-principles}
	\author{Ashkan Shekaari}
	\email{shekaari@email.kntu.ac.ir}
	\email{shekaari.theory@gmail.com}
	\author{Mahmoud Jafari}
	\author{Sana Afshar}
		\author{Rahil Mansouri}
	\affiliation{Department of Physics, K. N. Toosi University of Technology, Tehran 15875-4416, Iran}
	\date{\today}
	\begin{abstract}
The structural, electronic, and optical properties of monolayer $\alpha$-graphdiyne ($\alpha$-GDY) are systematically investigated using density-functional theory within the plane-wave pseudopotential formalism. The electronic band structure reveals a gapless Dirac crossing at the K point, indicating Dirac-semimetallic behavior within the PBE/GGA framework. The calculated total and orbital-projected density of states show that the electronic states near the Fermi level are dominated by the carbon $2p$ orbitals, while the contribution of the $2s$ orbitals is comparatively weak. The optical response exhibits pronounced polarization dependence. The in-plane dielectric function displays a strong low-energy electronic response and negative values of its real part, whereas the out-of-plane component remains positive throughout the investigated energy range. Consistently, the absorption coefficient, extinction coefficient, reflectivity, and electron energy-loss spectra reveal pronounced optical anisotropy. The calculated plasma frequencies are approximately $3.21$~eV for in-plane polarization and $1.06$~eV for out-of-plane polarization, highlighting the strongly anisotropic electronic response of the monolayer. These findings demonstrate that $\alpha$-GDY combines Dirac-like electronic behavior with highly anisotropic optical properties, indicating its potential relevance to polarization-sensitive optoelectronic, plasmonic, and nanoelectronic applications.
	\end{abstract}
		\keywords{$\alpha$-graphdiyne, 2D materials, Density-functional theory, Electronic structure, Optical anisotropy} 
	\maketitle
	\section{Introduction}
		Following the groundbreaking discovery of graphene~\cite{Novo}, low-dimensional materials have attracted enormous research interest owing to their exceptional properties~\cite{shek}.The remarkable success of graphene across diverse fields, ranging from materials science to nanomedicine~\cite{15,16,17}, has stimulated extensive efforts to discover and investigate new low-dimensional materials, including MoS$_2$~\cite{18}, TiC~\cite{19}, molybdenene~\cite{190}, phagraphene~\cite{pg2}, SLSiN~\cite{20}, and so on.
		
		Unlike conventional carbon allotropes, such as graphite, graphene, carbon nanotubes, and fullerenes, which are composed exclusively of sp$^2$-hybridized carbon atoms, carbon materials containing mixed sp and sp$^2$ hybridization provide a broader platform for exploring the structural diversity and electronic properties of carbon-based systems. Graphdiyne (GDY)~\cite{gdy} represents a prominent member of this class of carbon allotropes. It was theoretically proposed in 1997~\cite{ff} and was experimentally synthesized in 2010 as a 2D planar carbon network consisting of benzene rings (sp$^2$-C) interconnected by diacetylenic linkages (sp-C) on a copper substrate, exhibiting semiconducting behavior~\cite{gdy2}. Owing to its highly ordered carbon framework and unique structural characteristics, GDY is regarded as one of the most promising carbon allotropes~\cite{Z,X,C,V,B,N,A}. Consequently, it has attracted considerable attention for applications in heterogeneous catalysis~\cite{15a,16a,17a}, energy storage~\cite{18a,19a}, energy conversion~\cite{20a,21a,22a}, life sciences~\cite{23a}, and a wide variety of next-generation high-performance devices.
		
		GDY can be regarded as a graphene-derived carbon allotrope in which selected C--C bonds are replaced by diacetylenic linkages. Depending on the arrangement and concentration of these linkages, GDY is classified into several structural polymorphs, including $\alpha$-, $\beta$-, $\gamma$-, and 6,6,18-GDY~\cite{is}. Although considerable theoretical and experimental efforts have been devoted to GDY-based materials, comprehensive first-principles studies of the electronic and optical properties of $\alpha$-GDY remain relatively scarce. In particular, a systematic investigation combining its electronic structure and optical response within a unified computational framework is still lacking.
		
		Density-functional theory (DFT)~\cite{dft} has proven to be a reliable and widely adopted approach for describing the ground-state electronic properties of 2D carbon materials. Furthermore, frequency-dependent optical properties can be obtained from the complex dielectric function within the independent-particle approximation, providing valuable insight into light--matter interactions. Motivated by the limited theoretical understanding of $\alpha$-GDY, the present work employs first-principles DFT calculations~\cite{koh,mtn} to systematically investigate its electronic structure and optical properties, with particular emphasis on the band structure, density of states (DOS), dielectric function, refractive index, extinction coefficient, absorption coefficient, reflectivity, and electron energy-loss spectrum (EELS)~\cite{optik}.
		
		The remainder of this paper is organized as follows: Sec. II outlines the computational details; Sec. III presents and discusses the calculated structural, electronic, and optical properties; and finally, the conclusions are summarized in Section IV.
	\section{Computational Details}
		Self-consistent DFT calculations were performed within the plane-wave pseudopotential formalism~\cite{411,421} using the Perdew--Burke--Ernzerhof (PBE) exchange--correlation functional in the generalized gradient approximation (GGA)~\cite{431}, as implemented in the \qe\ package~\cite{451,qe2,qe3}. The interactions between the valence electrons and ionic cores were described using both scalar-relativistic ultrasoft pseudopotentials~\cite{461,471}, generated according to the Rappe--Rabe--Kaxiras--Joannopoulos (RRKJ) scheme~\cite{491} with nonlinear core correction~\cite{501}, and optimized norm-conserving Vanderbilt pseudopotentials~\cite{5000}. The valence electronic configuration of carbon was taken as $2s^22p^2$. The charge density was represented using the real-space Fourier interpolation scheme~\cite{511}. Plane-wave kinetic-energy cutoffs of 1088 and 8708~eV were employed for the wavefunctions and charge density, respectively. A vacuum layer exceeding 17~\AA\ was introduced along the $z$ direction to eliminate spurious interactions between periodically repeated images.
		
		The equilibrium lattice constant and atomic positions were obtained through both variable-cell and fixed-cell structural relaxations using the Broyden--Fletcher--Goldfarb--Shanno (BFGS) optimization algorithm~\cite{521,531}, together with variable-cell optimization based on the Murnaghan equation of state~\cite{4155,4255}. Structural optimization was continued until the residual Hellmann--Feynman forces~\cite{hf2,hf} on all atoms became smaller than $10^{-3}$~eV\,.\AA$^{-1}$.
		
		Brillouin-zone integrations were carried out using the Monkhorst--Pack $k$-point sampling scheme~\cite{mpk}, while partial occupancies were treated within the Methfessel--Paxton smearing method~\cite{541}. Electronic band structures were calculated along the high-symmetry $\Gamma$--M--K--$\Gamma$ path of the first Brillouin zone. All computational parameters were systematically optimized through energy-convergence tests to ensure the accuracy and reliability of the calculated results.
		
		It should be noted that the electronic and optical properties were obtained within the PBE/GGA as a routinely applied and computationally efficient approximating framework. The quantitative position of band crossings and electronic transition energies may be affected by the known limitations of semilocal exchange-correlation functionals. Higher-level approaches, such as HSE06 or GW, would be desirable for further validation of the band-crossing characteristics and optical transition energies of $\alpha$-GDY. 
		
		The optical response was also obtained within the independent-particle approximation and therefore does not explicitly account for electron--hole interactions. In 2D materials, the reduced dielectric screening can lead to strong excitonic effects, which may significantly modify the optical absorption spectrum. Consequently, excitonic interactions may alter the positions and intensities of the absorption features obtained here. A quantitative description of these effects would require a many-body treatment, such as GW--BSE calculations, which is beyond the computational scope of the present study. The present optical results should therefore be interpreted as the independent-particle optical response of $\alpha$-GDY.
		
		\section{Results and discussion}
		\subsection{Equilibrium structural parameters}
		The equilibrium structural parameters of $\alpha$-GDY were determined through a combination of variable-cell and fixed-cell structural relaxations using the BFGS optimization algorithm, together with variable-cell optimization based on the Murnaghan equation of state. The optimized lattice constant and atomic coordinates are summarized in Table~\ref{tab:1}, while the corresponding relaxed crystal structure is illustrated in Fig.~\ref{fig:1}.
				\begin{widetext}
			\begin{minipage}{\linewidth} 
		\begin{figure}[H]
	\centering
	\subfigure[]{\label{subfig:1(a)}
		\includegraphics[scale=0.03]{fig1-a.pdf}}
	\subfigure[]{\label{subfig:1(b)}
		\includegraphics[scale=0.04]{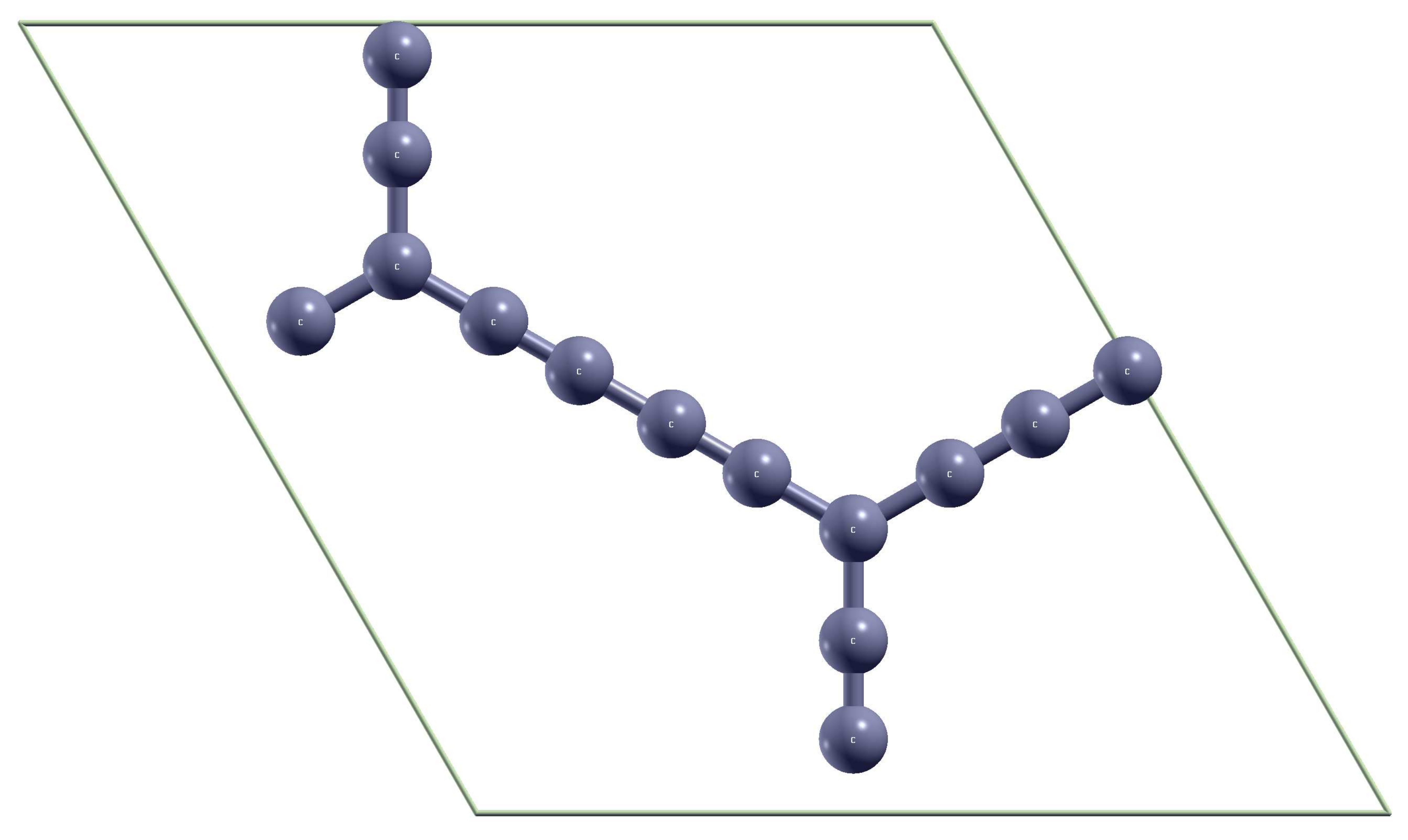}}
	\caption{\label{fig:1}
		Atomic structures of (a) a $4\times4$ supercell and (b) the unit cell of $\alpha$-GDY, rendered using XCrySDen~\cite{xcr}. The hexagonal unit cell contains 14 carbon atoms. The boundary of each hexagonal pore consists of 30 carbon atoms, with 5 atoms along each side.}
\end{figure}		
\end{minipage}
\end{widetext}
		The equilibrium lattice constant was further determined by fitting the calculated energy--volume data to the Murnaghan equation of state, which assumes a linear pressure dependence of the bulk modulus under finite strain. Following structural relaxation, a series of uniformly strained configurations was generated by varying the lattice constant in increments of approximately $\pm2.65\times10^{-2}$~\AA, resulting in nine distinct unit-cell volumes.
		\begin{table}[H]
	\caption{\label{tab:1}
		Relaxed atomic structure of the $\alpha$-GDY unit cell, showing the Cartesian coordinates of the 14 carbon atoms projected onto the $x$--$y$ plane.}
	\begin{tabular}{cl}
		\hline
		$x$ (\AA) & $ $ $y$ (\AA)\\
		\hline		
		-3.2672997156  $ $&  5.5319469889\\
		4.7109885423  $ $&  0.9257882246\\
		6.9875281113  $ $&  4.8671708500\\
		-0.9921406341  $ $&  9.4742899952\\
		4.7111936359  $ $&  2.1608172000\\
		-0.9917307841  $ $&  6.8464033190\\
		0.2146167189  $ $&  6.1502793515\\
		-2.1977477756  $ $&  6.1497541942\\
		3.5053629381  $ $&  4.2503630789\\
		2.4359217023  $ $&  4.8679572996\\
		-0.9920514460  $ $&  8.2392454021\\
		1.2843317104  $ $&  5.5328289377\\
		5.9179135618  $ $&  4.2497990237\\
		4.7114913533  $ $&  3.5536478274\\
		\hline
	\end{tabular}
\end{table}
		
		Self-consistent field calculations were subsequently performed for each strained configuration to obtain the corresponding total energies. A least-squares fit of the calculated energy--volume data to the Murnaghan equation of state yielded an equilibrium lattice constant of $11.4064$~\AA, in good agreement with previously reported theoretical results.
		\subsection{Electronic structure}
		\subsubsection{Band structure and total DOS}
		Figure~\ref{subfig:2(a)} presents the calculated electronic band structure of $\alpha$-GDY along the high-symmetry $\Gamma-$M$-$K$-\Gamma$ path of the first Brillouin zone. The most prominent feature is that the valence band (VB) and conduction band (CB) intersect exactly at the K point without opening an energy gap, forming a Dirac point. Consequently, the band gap is zero ($E_g=0$ eV), indicating the metallic (Dirac semimetallic) nature of the monolayer~\cite{fct}. In the vicinity of the K point, the two bands exhibit an almost linear energy dispersion, which is a characteristic signature of massless Dirac fermions~\cite{mdf}. The Fermi level passes directly through the band-crossing point, confirming that no forbidden energy gap exists between the occupied and unoccupied states. This observation is fully consistent with the calculated total DOS, which exhibits a finite electronic density at the Fermi energy, further supporting the conducting character of the material.
						\begin{widetext}
			\begin{minipage}{\linewidth} 
		\begin{figure}[H]
			\centering
	\subfigure[]{\label{subfig:2(a)}
	\includegraphics[scale=0.44]{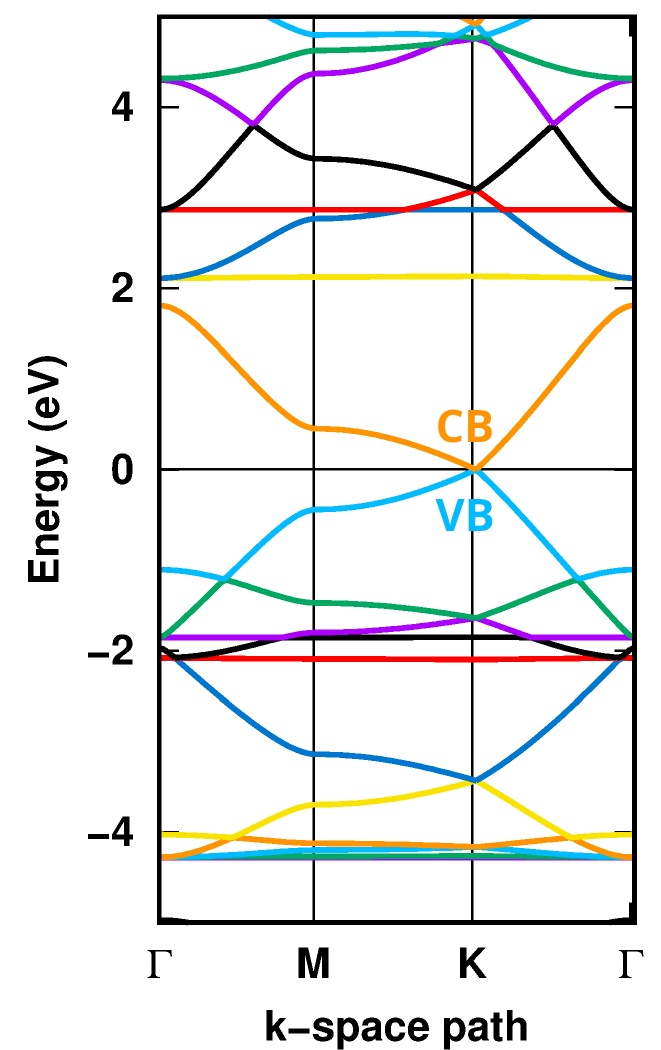}}
\subfigure[]{\label{subfig:2(b)}
	\includegraphics[scale=0.3]{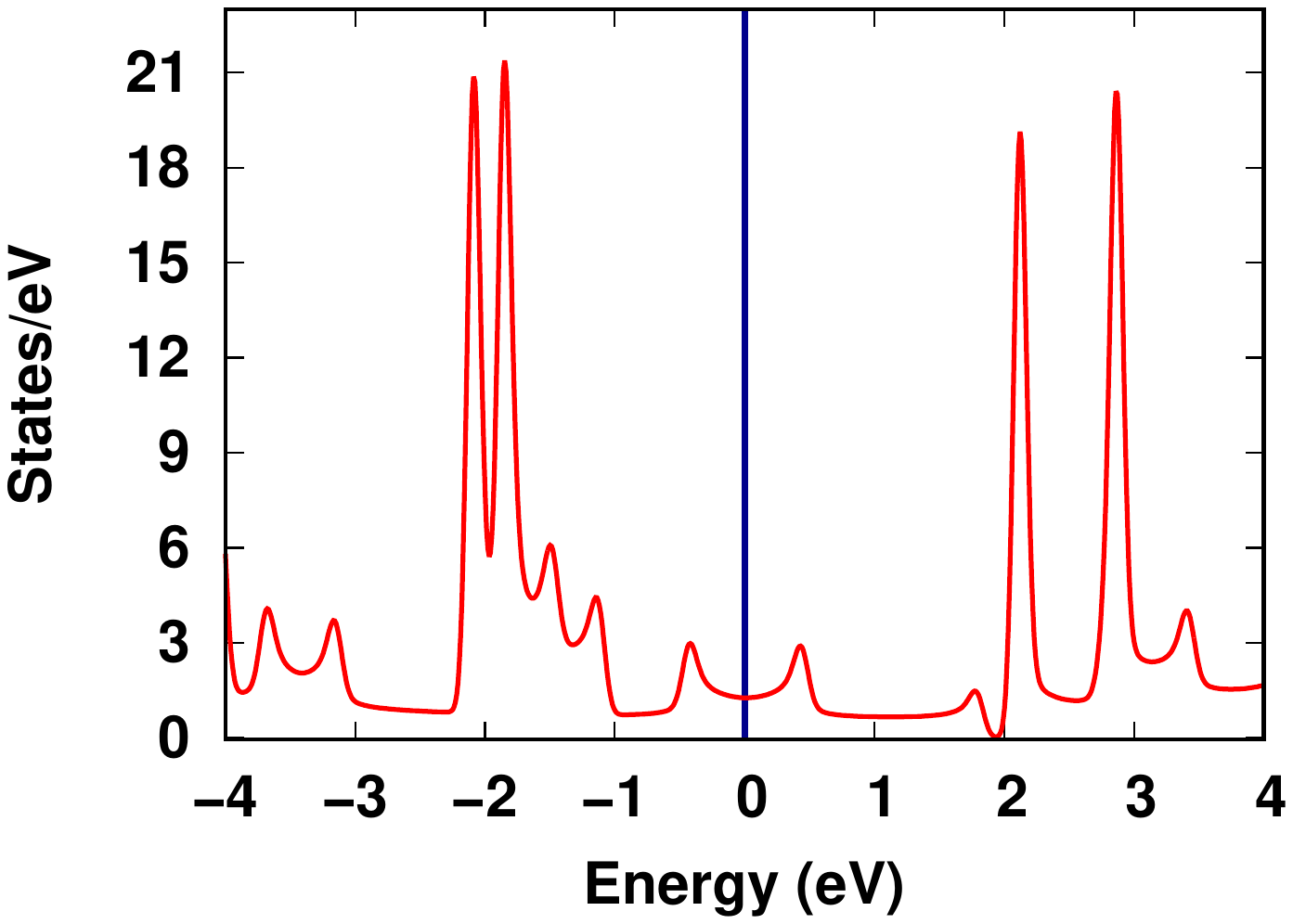}}
			\caption{\label{fig:2}
				Electronic band structure (a), and total DOS (b) of $\alpha$-GDY. The valence (VB) and conduction (CB) bands meet at the K point, forming a Dirac point similar to that in graphene, which confirms the metallic character of the monolayer. The Fermi level is set to 0~eV. The nonzero DOS at the Fermi level confirms the metallic character of the system, consistent with the band structure.}
		\end{figure}
	\end{minipage}
\end{widetext}
		
		Away from the K point, the electronic bands become more dispersive, particularly along the $\Gamma-$M and K$-\Gamma$ directions, indicating significant orbital hybridization and electron delocalization throughout the carbon network. The noticeable variation of the band curvature along different crystallographic directions also reflects the anisotropic electronic response of the monolayer.
		
		Overall, the calculated band structure demonstrates that $\alpha$-GDY is a gapless 2D carbon material with a graphene-like Dirac crossing at the K point. Such an electronic structure is highly desirable for high-speed electronic devices, nanoelectronics, and optoelectronic applications because it combines metallic conductivity with the unique transport properties associated with Dirac fermions.
		
		Figure~\ref{subfig:2(b)} shows the calculated total DOS of $\alpha$-GDY, with the Fermi level set at 0 eV. A finite DOS is observed at the Fermi level, indicating that the system possesses a metallic electronic character. This result is fully consistent with the electronic band structure, where the valence and conduction bands touch at the K point without opening a band gap.
		
		Several pronounced DOS peaks are observed in both the valence and conduction regions, which correspond to Van Hove singularities~\cite{van} originating from relatively flat electronic bands. In the valence band, the most intense peaks are located around $-2$ eV, whereas similar sharp features appear in the conduction band near $2-3$ eV. These peaks indicate energy ranges with a high concentration of electronic states and are expected to contribute significantly to interband optical transitions.
		
		In contrast, the DOS is considerably lower in the vicinity of the Fermi level than at the neighboring peaks, implying that only a limited number of electronic states participate in low-energy excitations. Nevertheless, the nonzero DOS at the Fermi energy confirms the absence of a forbidden energy gap and supports the metallic transport behavior predicted by the band-structure calculations.
		
		The DOS exhibits an approximately symmetric distribution of occupied and unoccupied states around the Fermi level, although the peak intensities are not identical on the two sides. Such a feature is characteristic of carbon-based 2D Dirac materials~\cite{dir} and reflects the electronic structure governed predominantly by the hybridized carbon $2p$ orbitals.
		
		The combination of a finite DOS at the Fermi level and pronounced Van Hove singularities away from $E_F$ suggest that $\alpha$-GDY may exhibit efficient electronic transport while simultaneously providing strong optical transitions at higher photon energies, making it a promising candidate for nanoelectronic and optoelectronic applications.\\
		
		\subsubsection{Orbital-projected density of states (PDOS)}
		Figure~\ref{fig:3} presents the PDOS of $\alpha$-GDY for the $s$, $p_x$, $p_y$, and $p_z$ orbitals. The results reveal that the electronic structure is dominated by the carbon $2p$ states, whereas the contribution of the $2s$ orbitals is comparatively small over the entire energy range. The $2s$ orbitals exhibit relatively weak and localized peaks in both the valence and conduction bands, indicating that they contribute only marginally to the electronic states near the Fermi level. Consequently, the $2s$ states play a limited role in determining the electrical conductivity and low-energy electronic excitations of the monolayer.
		\begin{widetext}
		\begin{minipage}{\linewidth} 
		\begin{figure}[H]
			\centering
	\subfigure[]{\label{subfig:3(a)}
	\includegraphics[scale=0.31]{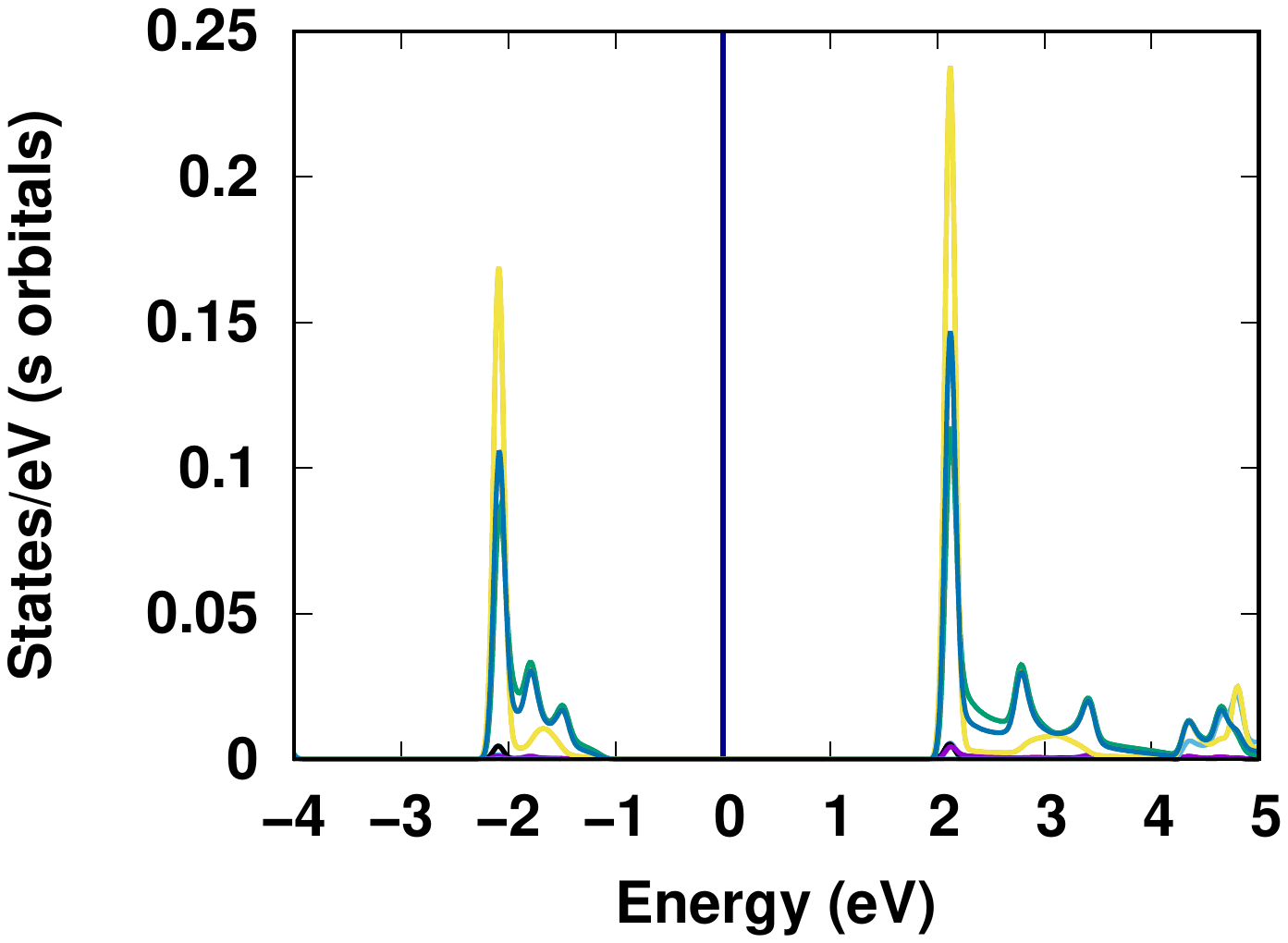}}
\subfigure[]{\label{subfig:3(b)}
	\includegraphics[scale=0.31]{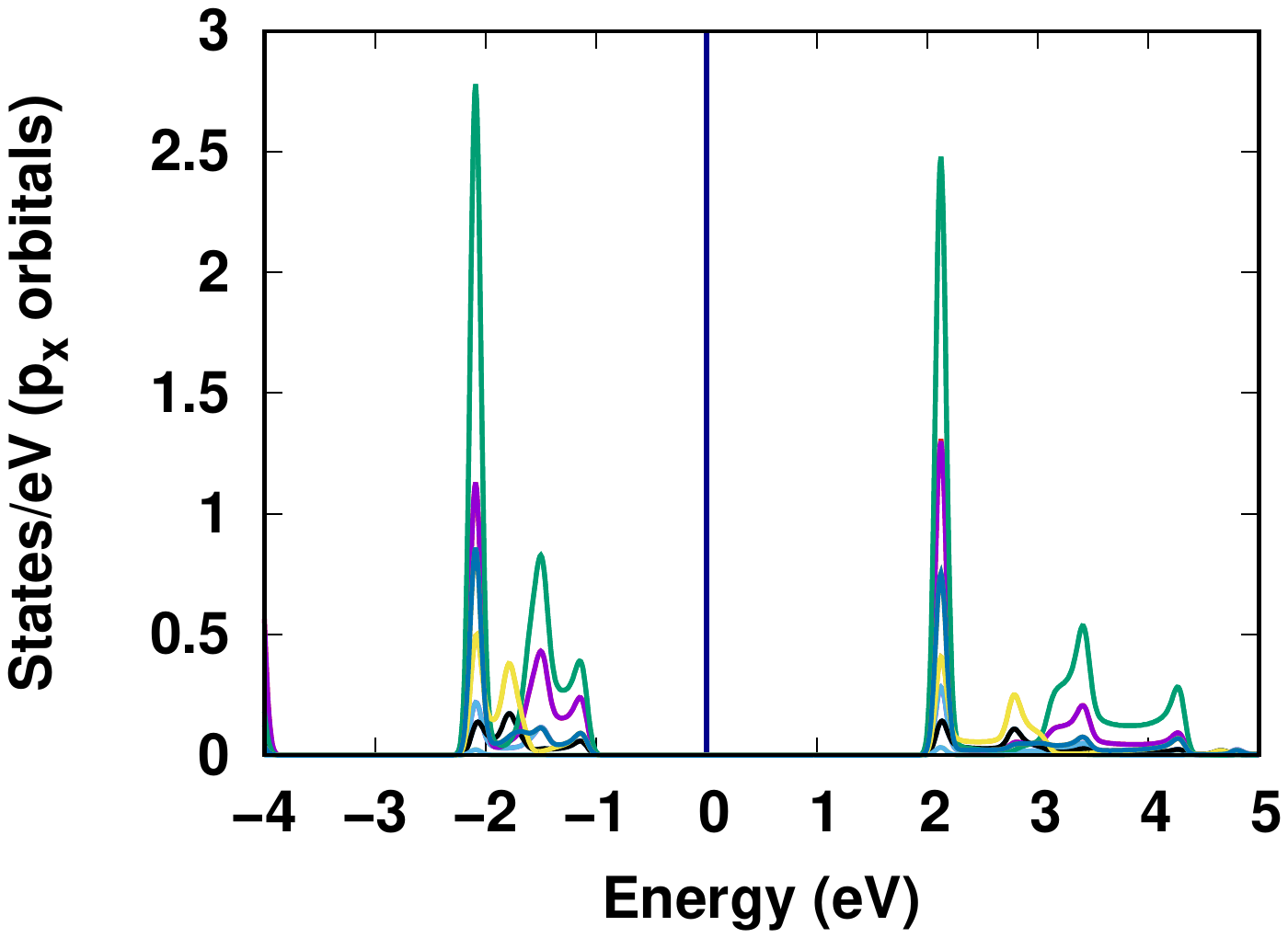}}
			\end{figure}
		\end{minipage}
\end{widetext}
		\begin{widetext}
	\begin{minipage}{\linewidth} 
		\begin{figure}[H]
			\centering
		\subfigure[]{\label{subfig:3(c)}
		\includegraphics[scale=0.31]{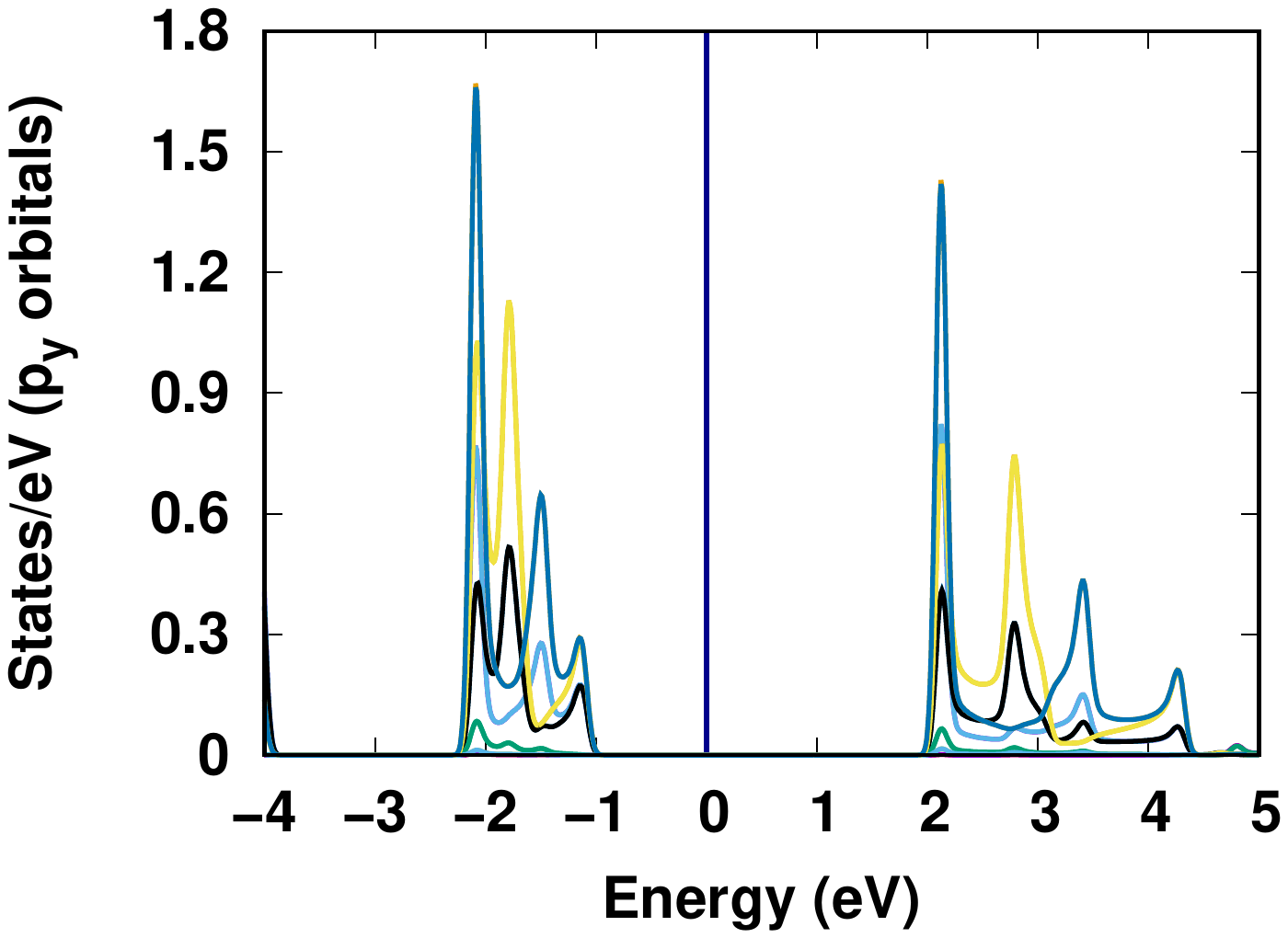}}
	\subfigure[]{\label{subfig:3(d)}
		\includegraphics[scale=0.31]{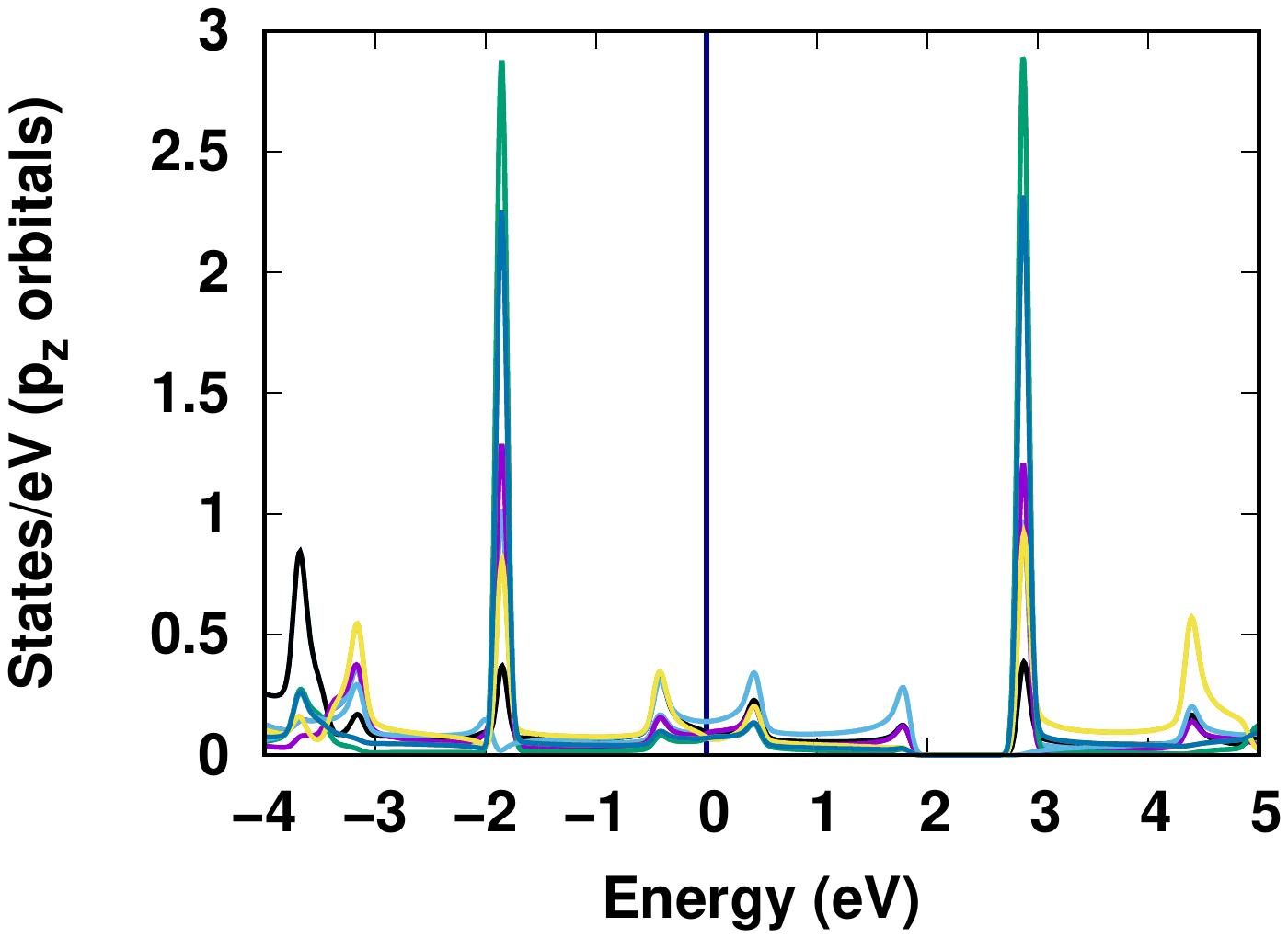}}
			\caption{\label{fig:3}
				Calculated orbital-projected DOS of $\alpha$-GDY. The contributions from the (a) $s$, (b) $p_x$, (c) $p_y$, and (d) $p_z$ orbitals are presented to reveal their roles in the electronic structure. The Fermi level is set to 0~eV.}
		\end{figure}
	\end{minipage}
\end{widetext}
		
		In contrast, the $2p_x$, $2p_y$, and $2p_z$ orbitals contribute significantly to the electronic structure. The $p_x$ and $p_z$ orbitals exhibit the largest PDOS peaks, particularly around $-2$ eV and 2 eV, demonstrating that these orbitals dominate the occupied and unoccupied electronic states. The $p_y$ orbital also contributes substantially but with comparatively smaller peak intensities. Near the Fermi level, the DOS is primarily composed of the carbon 2p orbitals, while the contribution of the $2s$ orbital remains negligible. This observation indicates that the metallic (Dirac semimetallic) character of $\alpha$-GDY originates predominantly from the hybridized carbon $2p$ electronic states.
		
		The sharp PDOS peaks observed away from the Fermi level correspond to Van Hove singularities arising from relatively flat electronic bands. The coincidence of peak positions among the $p_x$, $p_y$, and $p_z$ orbitals suggests significant orbital hybridization throughout the carbon network. Such hybridization is expected to enhance electron delocalization and plays an essential role in both the electronic transport and optical transitions of the material.
		
		Although the major peaks of the $p_x$, $p_y$, and $p_z$ orbitals appear at similar energies, their intensities differ considerably. In particular, the $p_x$ and $p_z$ orbitals exhibit stronger contributions than the $p_y$ orbital over most of the investigated energy window, indicating an anisotropic distribution of the electronic states among the $2p$ orbitals. Overall, the PDOS analysis demonstrates that the low-energy electronic properties of $\alpha$-GDY are governed almost entirely by the carbon $2p$ orbitals, whereas the $2s$ states mainly contribute to deeper valence and higher conduction states. This orbital distribution is consistent with the calculated band structure and explains the metallic electronic behavior of the monolayer.
		\subsection{Optical properties}
		\subsubsection{Real part of the dielectric function}
		Figure~\ref{fig:4} presents the calculated real part of the dielectric function, $\varepsilon_1(\omega)$, of $\alpha$-GDY for the in-plane and out-of-plane polarization directions. A pronounced optical anisotropy is observed between the two components over the entire investigated energy range. The in-plane dielectric response exhibits strong oscillations in the low-energy region, including several positive and negative excursions below approximately $2$~eV.

		The negative values of $\varepsilon_1(\omega)$ indicate metallic behavior within specific energy intervals, where the incident electromagnetic wave is efficiently screened by free charge carriers. As the photon energy increases, these oscillations gradually diminish, and $\varepsilon_1(\omega)$ approaches a nearly constant value close to unity above approximately $6$~eV, indicating a reduced dielectric response at higher photon energies.

		In contrast, the out-of-plane component remains positive throughout the investigated energy range and varies smoothly without pronounced resonances. Starting from a static dielectric constant of approximately $1.1$, $\varepsilon_1^{z}(\omega)$ increases gradually with photon energy and reaches a value of about $1.5$ near $10$~eV. Thus, $\varepsilon_1^{z}(\omega)$ remains only slightly larger than the vacuum value of unity, with an increase of approximately $0.4$ over the investigated range. This relatively small magnitude indicates a weak out-of-plane electronic polarizability and, consequently, weaker electronic screening compared with the in-plane response. The absence of negative values further indicates that no strong out-of-plane plasmon-like response is obtained within the investigated energy range.
						\begin{widetext}
			\begin{minipage}{\linewidth} 
				\begin{figure}[H]
					\centering
					\subfigure[]{\label{subfig:4(a)}
						\includegraphics[scale=0.31]{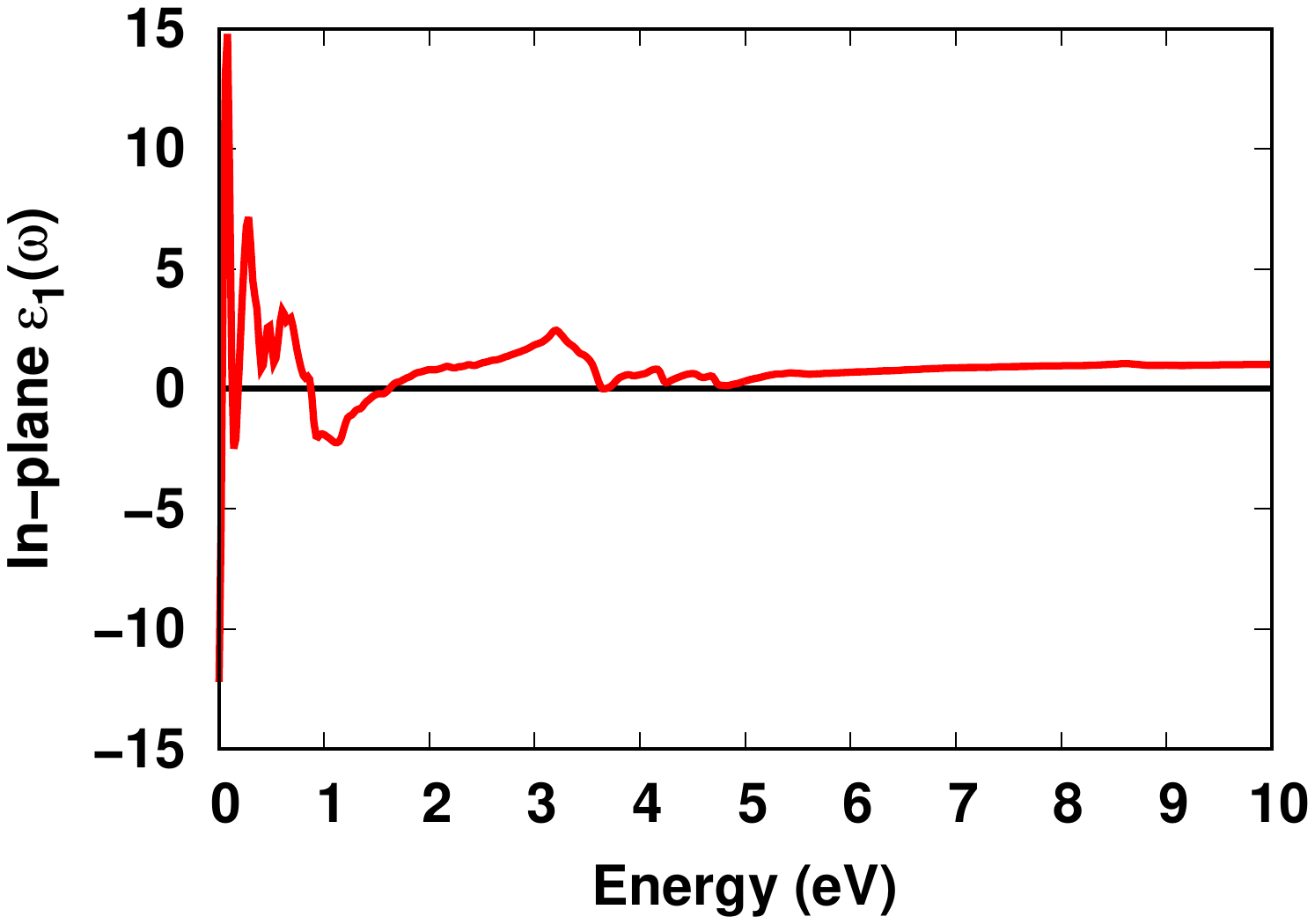}}
					\subfigure[]{\label{subfig:4(b)}
						\includegraphics[scale=0.31]{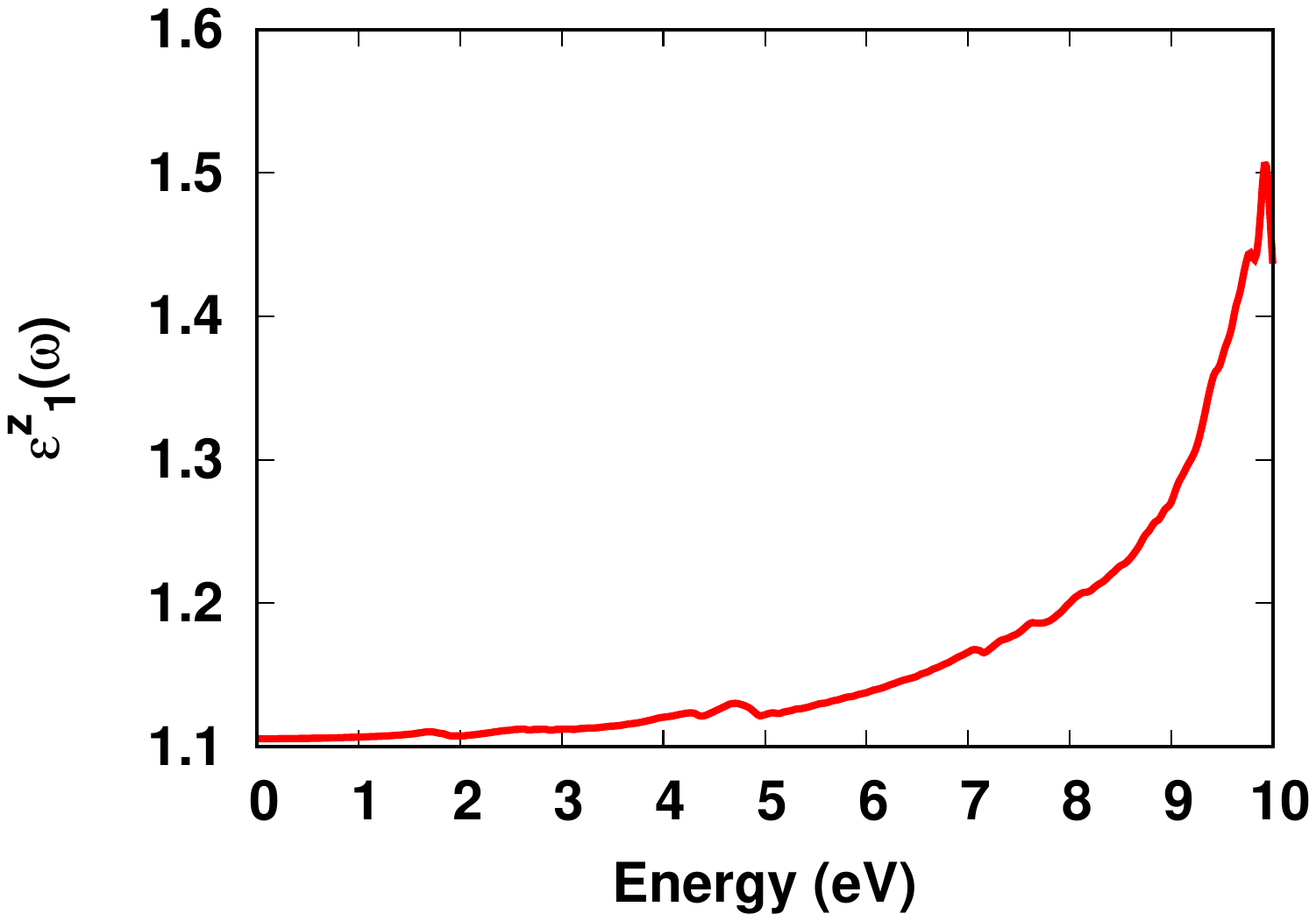}}
					\caption{\label{fig:4}
						Calculated real part of the dielectric function, $\epsilon_1(\omega)$, as a function of photon energy for $\alpha$-GDY. The (a) in-plane ($x-y$ plane) and (b) out-of-plane ($z$ direction, perpendicular to the monolayer plane) components are presented to illustrate the optical anisotropy.}
				\end{figure}
			\end{minipage}
		\end{widetext}
		The distinct difference between the in-plane and out-of-plane dielectric responses reflects the intrinsic optical anisotropy of the $\alpha$-GDY monolayer. This behavior originates from the strong covalent bonding and highly delocalized $\pi$ electrons within the atomic plane, whereas the electronic polarization perpendicular to the monolayer is substantially weaker, leading to a much smaller dielectric response.
		\subsubsection{Imaginary part of the dielectric function}
		Figure~\ref{fig:5} presents the calculated imaginary part of the dielectric function, $\varepsilon_2(\omega)$, of $\alpha$-GDY for the in-plane and out-of-plane polarization directions. A pronounced optical anisotropy is observed between the two polarization directions. The in-plane component exhibits an extremely intense peak in the low-energy region, which is characteristic of the Drude response arising from free charge carriers in metallic systems. This strong low-energy dielectric loss confirms the metallic character of the monolayer and is fully consistent with the band-structure and DOS analyses.
		
		Following the Drude peak~\cite{drp}, $\varepsilon_2(\omega)$ decreases rapidly with increasing photon energy and remains relatively small over the rest of the investigated energy range, indicating a reduced contribution of free-carrier absorption at higher photon energies.
		
		In contrast, the out-of-plane component remains nearly zero over most of the investigated energy range and increases only gradually above approximately $8$~eV. Even at its maximum, the magnitude of $\varepsilon_2^{z}(\omega)$ is several orders of magnitude smaller than that of the in-plane component, demonstrating that dielectric losses are strongly suppressed for light polarized perpendicular to the monolayer.
		
		The marked difference between the in-plane and out-of-plane dielectric responses highlights the intrinsic optical anisotropy of $\alpha$-GDY. The strong in-plane dielectric loss originates from the highly delocalized $\pi$ electrons within the carbon network, whereas the weak out-of-plane response reflects the limited electronic polarization normal to the monolayer.
						\begin{widetext}
			\begin{minipage}{\linewidth}
				\begin{figure}[H]
					\centering
					\subfigure[]{\label{subfig:5(a)}
						\includegraphics[scale=0.31]{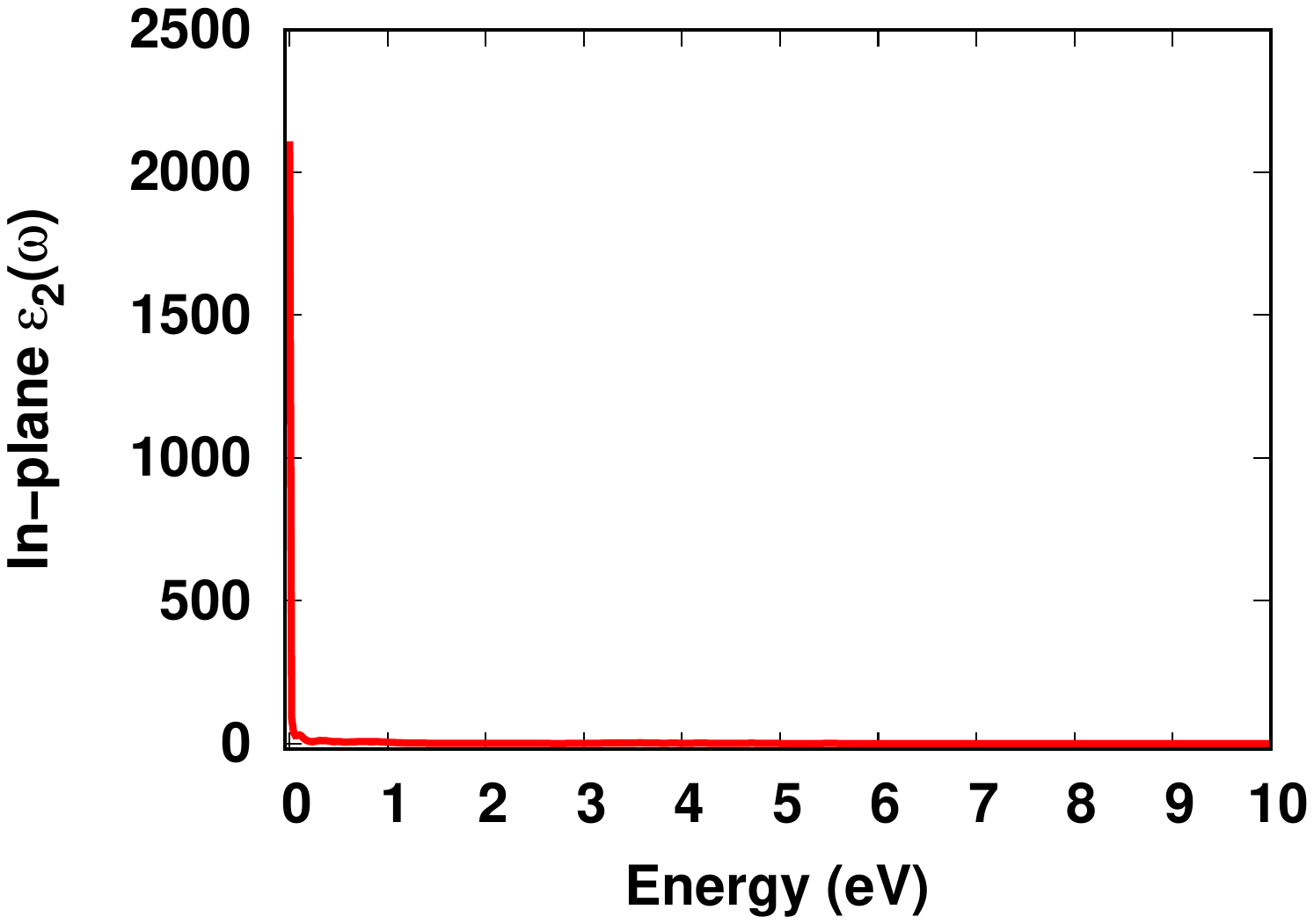}}
					\subfigure[]{\label{subfig:5(b)}
						\includegraphics[scale=0.31]{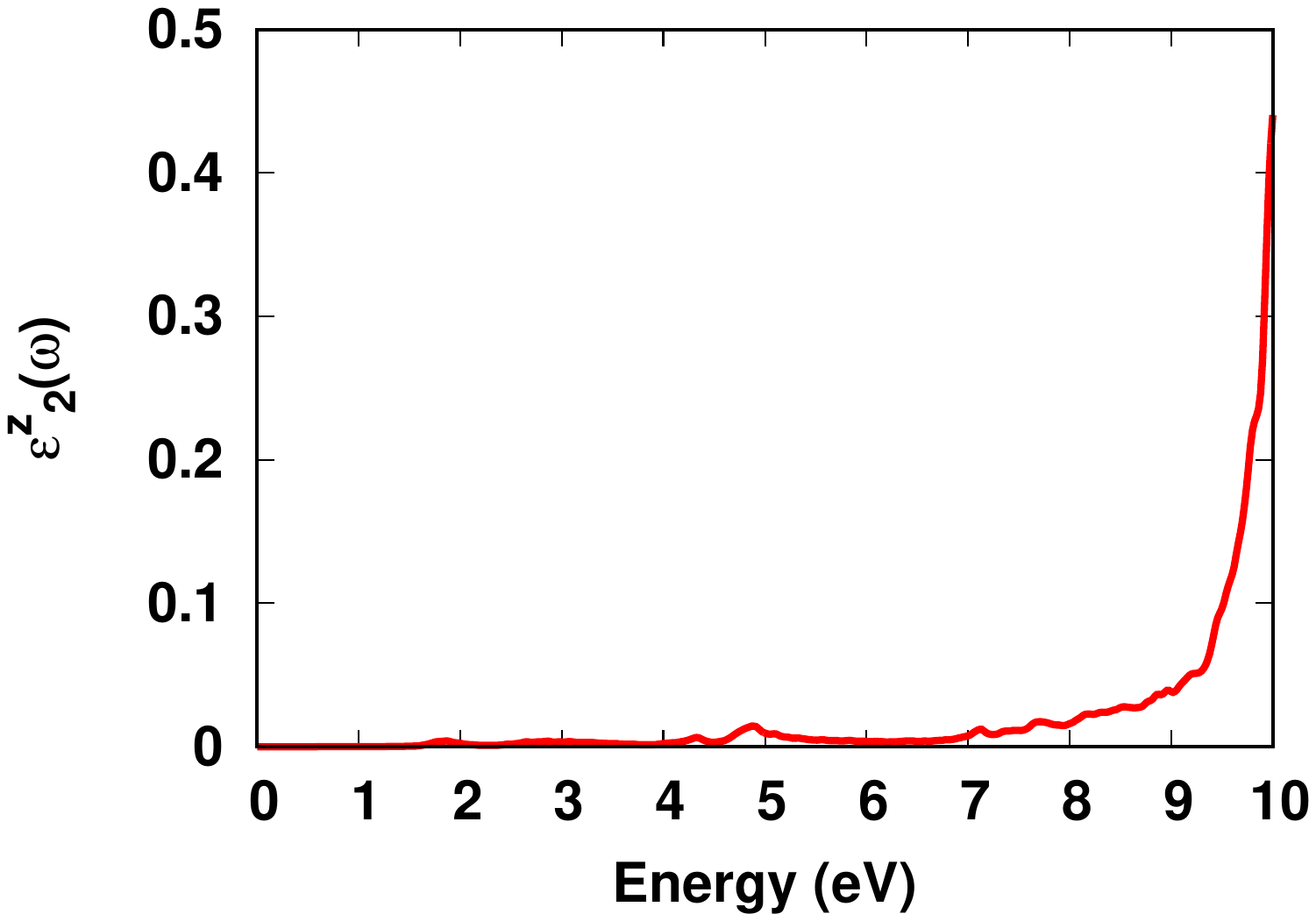}}
					\caption{\label{fig:5}
						Calculated imaginary part of the dielectric function, $\epsilon_2(\omega)$, as a function of photon energy for $\alpha$-GDY. The (a) in-plane ($x-y$ plane) and (b) out-of-plane ($z$ direction, perpendicular to the monolayer plane) components are shown for comparison.}
				\end{figure}
			\end{minipage}
		\end{widetext}
		\subsubsection{Absorption coefficient}
		Figure~\ref{fig:6} presents the calculated absorption coefficient, $\alpha(\omega)$, of $\alpha$-GDY for the in-plane and out-of-plane polarization directions. A pronounced optical anisotropy is observed throughout the investigated photon-energy range. The in-plane absorption coefficient exhibits several intense absorption bands extending from the infrared to the ultraviolet region. The absorption begins at nearly zero photon energy owing to the gapless electronic structure of $\alpha$-GDY and reaches its strongest intensity between approximately $3$ and $5$~eV, where the absorption coefficient exceeds $3.5\times10^{5}$~cm$^{-1}$. These intense absorption bands originate from the high probability of electronic transitions between occupied and unoccupied states, consistent with the prominent features observed in the imaginary part of the dielectric function.
				\begin{widetext}
			\begin{minipage}{\linewidth}
		\begin{figure}[H]
			\centering
	\subfigure[]{\label{subfig:6(a)}
	\includegraphics[scale=0.31]{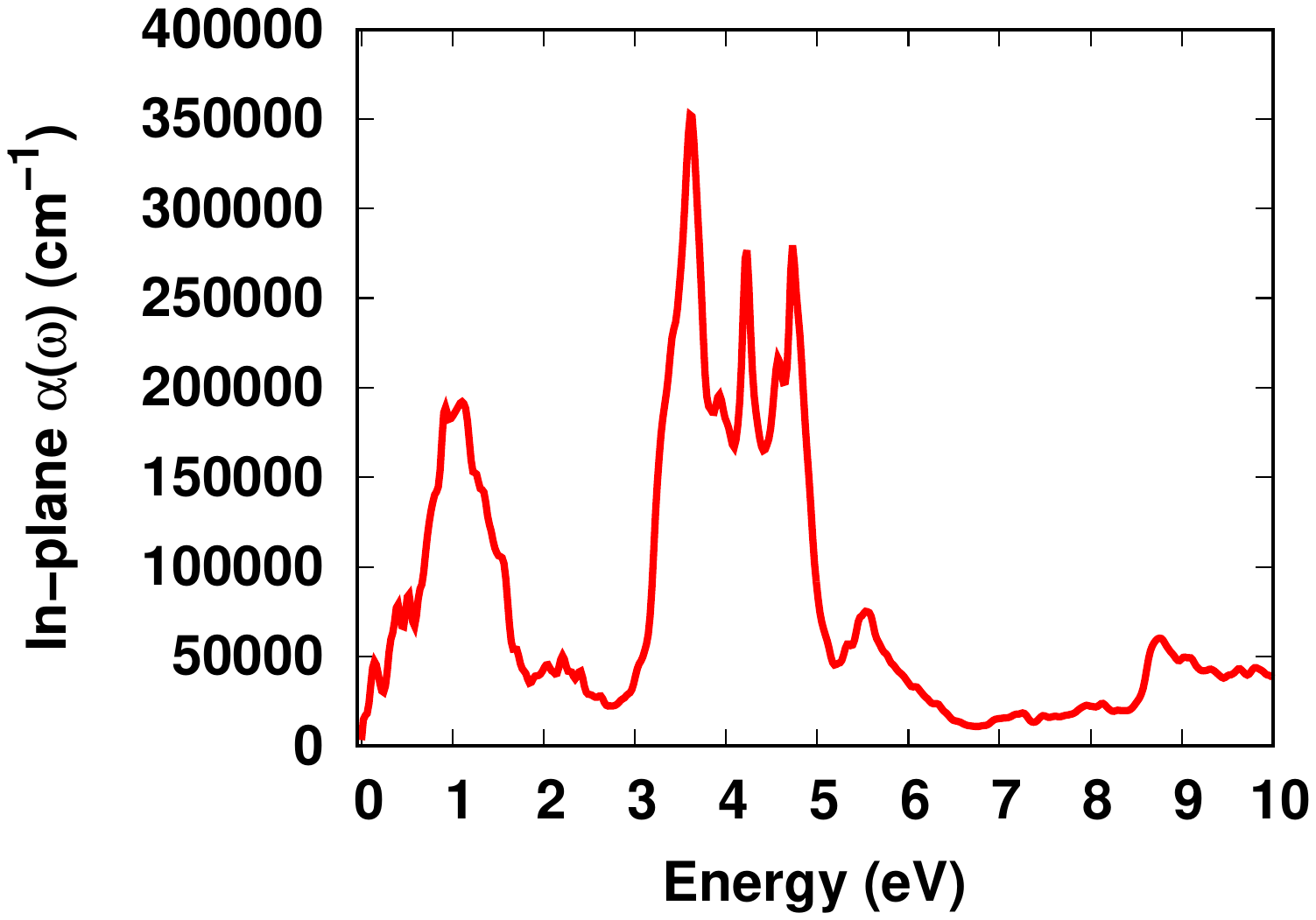}}
\subfigure[]{\label{subfig:6(b)}
	\includegraphics[scale=0.31]{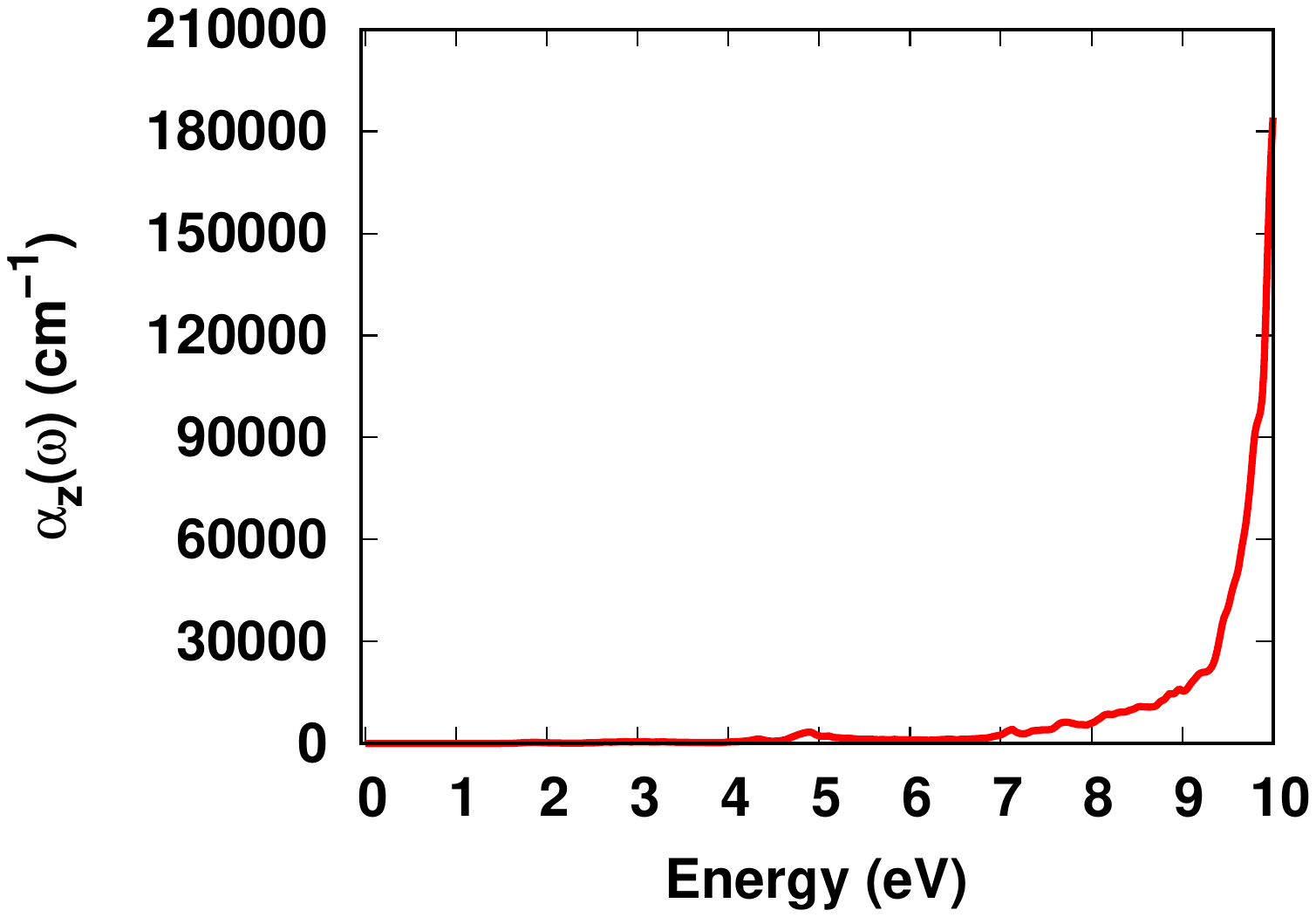}}
			\caption{\label{fig:6}
				Calculated absorption coefficient, $\alpha(\omega)$, as a function of photon energy for $\alpha$-GDY. The (a) in-plane ($x-y$ plane) and (b) out-of-plane ($z$ direction, perpendicular to the monolayer plane) components are presented to illustrate the optical anisotropy.}
		\end{figure}
	\end{minipage}
\end{widetext}
		Beyond the principal absorption region, the absorption coefficient decreases gradually with increasing photon energy while remaining finite over the entire investigated energy range. The persistence of nonzero absorption demonstrates that $\alpha$-GDY is capable of interacting efficiently with electromagnetic radiation over a broad spectral range.
		
		In contrast, the out-of-plane absorption, $\alpha_z(\omega)$, remains extremely weak below approximately $8$~eV and increases noticeably only at higher photon energies, reaching a maximum value close to $1.8\times10^{5}$~cm$^{-1}$ near $10$~eV. The substantially smaller absorption coefficient compared with the in-plane response indicates that optical transitions polarized perpendicular to the monolayer are strongly suppressed.
		
		The remarkable difference between the in-plane and out-of-plane absorption spectra confirms the intrinsic optical anisotropy of $\alpha$-GDY. The strong in-plane absorption originates from the highly delocalized $\pi$ electrons within the carbon framework, whereas the weak out-of-plane response reflects the limited electronic polarization normal to the atomic plane. Such highly anisotropic absorption behavior suggests that $\alpha$-GDY is a promising candidate for polarization-sensitive optoelectronic devices and broadband photonic applications.
		\subsubsection{Reflectivity}
		Figure~\ref{fig:7} presents the calculated reflectivity spectra, $R(\omega)$, of $\alpha$-GDY for both in-plane and out-of-plane polarization directions. A pronounced optical anisotropy is observed throughout the investigated photon-energy range. The in-plane reflectivity exhibits a very high value in the low-energy region, approaching unity as the photon energy tends toward zero. This behavior is characteristic of metallic systems and originates from the strong reflection of low-energy electromagnetic radiation by free charge carriers, consistent with the Drude-like response observed in the imaginary part of the dielectric function (FIG.~\ref{fig:6}).
						\begin{widetext}
			\begin{minipage}{\linewidth}
				\begin{figure}[H]
					\centering
					\subfigure[]{\label{subfig:7(a)}
						\includegraphics[scale=0.31]{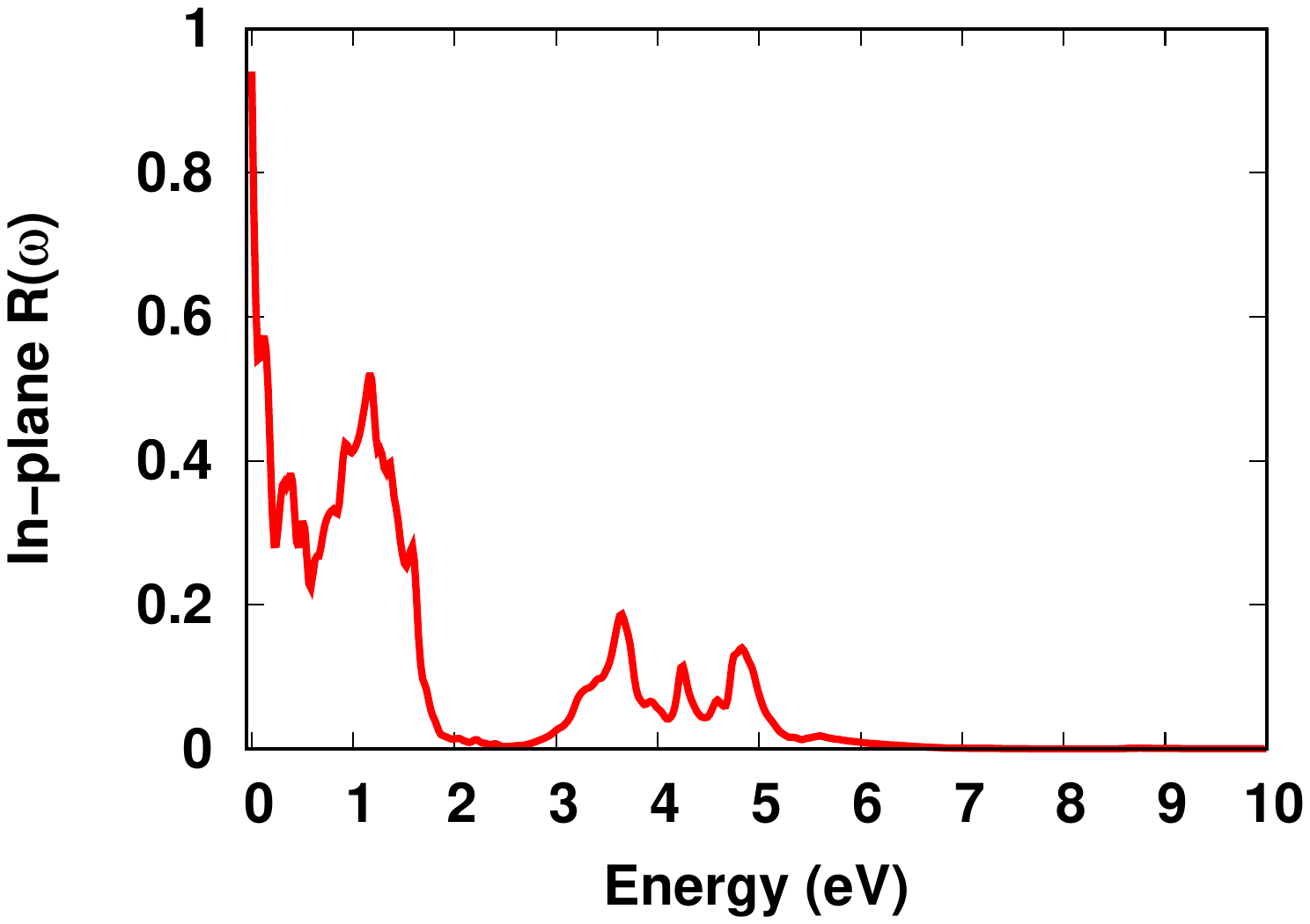}}
					\subfigure[]{\label{subfig:7(b)}
						\includegraphics[scale=0.31]{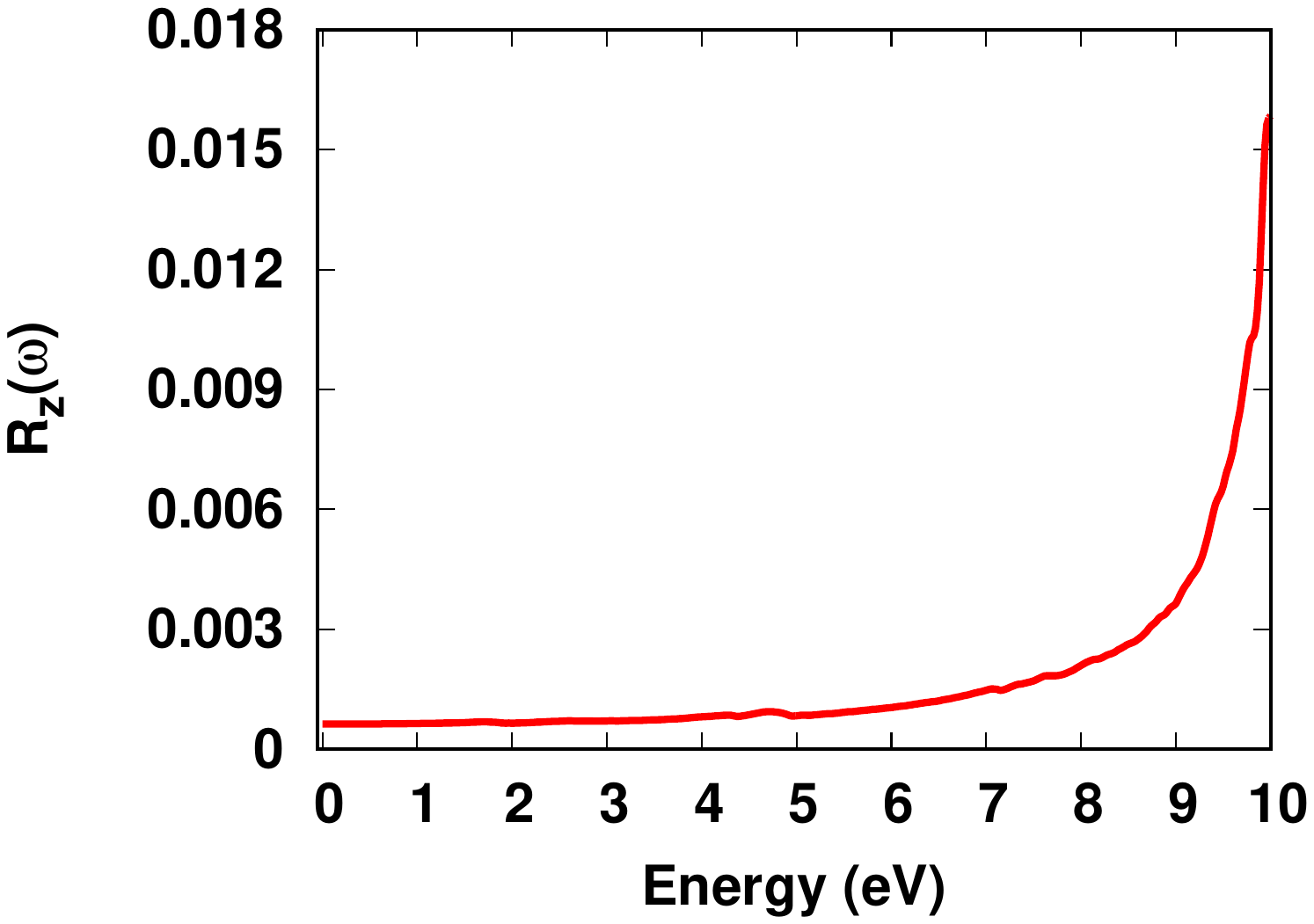}}
					\caption{\label{fig:7}
						Calculated reflectivity, $R(\omega)$, as a function of photon energy for $\alpha$-GDY. The (a) in-plane ($x-y$ plane) and (b) out-of-plane ($z$ direction, perpendicular to the monolayer plane) components are shown for comparison.}
				\end{figure}
			\end{minipage}
		\end{widetext}
		As the photon energy increases, the in-plane reflectivity decreases rapidly, accompanied by several broad and weak maxima between approximately $3$ and $5$~eV. These spectral features are closely correlated with the interband electronic transitions responsible for the peaks in the absorption coefficient and the imaginary part of the dielectric function. Above approximately $6$~eV, the reflectivity becomes negligibly small, indicating that most of the incident electromagnetic radiation is transmitted or absorbed rather than reflected.
		
		In contrast, the out-of-plane reflectivity, $R_z(\omega)$, remains extremely low over the entire investigated energy range, increasing only gradually above approximately $8$~eV and reaching a maximum value of less than $2\%$ near $10$~eV. Such weak reflection demonstrates that electromagnetic waves polarized perpendicular to the monolayer interact only weakly with the electronic system.
		
		The marked difference between the in-plane and out-of-plane reflectivity spectra further confirms the intrinsic optical anisotropy of $\alpha$-GDY. The strong in-plane reflection originates from the highly delocalized $\pi$ electrons within the carbon framework, whereas the weak out-of-plane response reflects the limited electronic polarization normal to the monolayer. These results suggest that $\alpha$-GDY exhibits highly polarization-dependent optical behavior, making it a promising material for anisotropic photonic and optoelectronic applications.\\
		\subsubsection{Extinction coefficient}
		Figure~\ref{fig:8} presents the calculated extinction coefficient, $k(\omega)$, of $\alpha$-GDY for the in-plane and out-of-plane polarization directions. The extinction spectra exhibit pronounced optical anisotropy throughout the investigated photon-energy range. The in-plane component displays a strong low-energy response, followed by a rapid decrease with increasing photon energy. This behavior reflects strong attenuation of electromagnetic radiation associated with the pronounced low-energy electronic response of $\alpha$-GDY and is consistent with the corresponding behavior of the imaginary part of the dielectric function (Fig.~\ref{fig:5}).

		As the photon energy increases, the in-plane extinction coefficient decreases markedly and exhibits relatively weak structures between approximately $3$ and $5$~eV, which are associated with electronic transitions contributing to the optical response. Above this energy range, $k(\omega)$ approaches nearly zero, indicating substantially weaker attenuation of electromagnetic radiation at higher photon energies.

		In contrast, the out-of-plane extinction coefficient remains almost negligible over most of the investigated energy range and increases gradually only above approximately $8$~eV, reaching its maximum value near $10$~eV. The maximum out-of-plane extinction coefficient is approximately one order of magnitude smaller than the corresponding in-plane response in the relevant energy range, demonstrating the substantially weaker attenuation of electromagnetic waves polarized perpendicular to the monolayer.
						\begin{widetext}
			\begin{minipage}{\linewidth}
				\begin{figure}[H]
					\centering
					\subfigure[]{\label{subfig:8(a)}
						\includegraphics[scale=0.31]{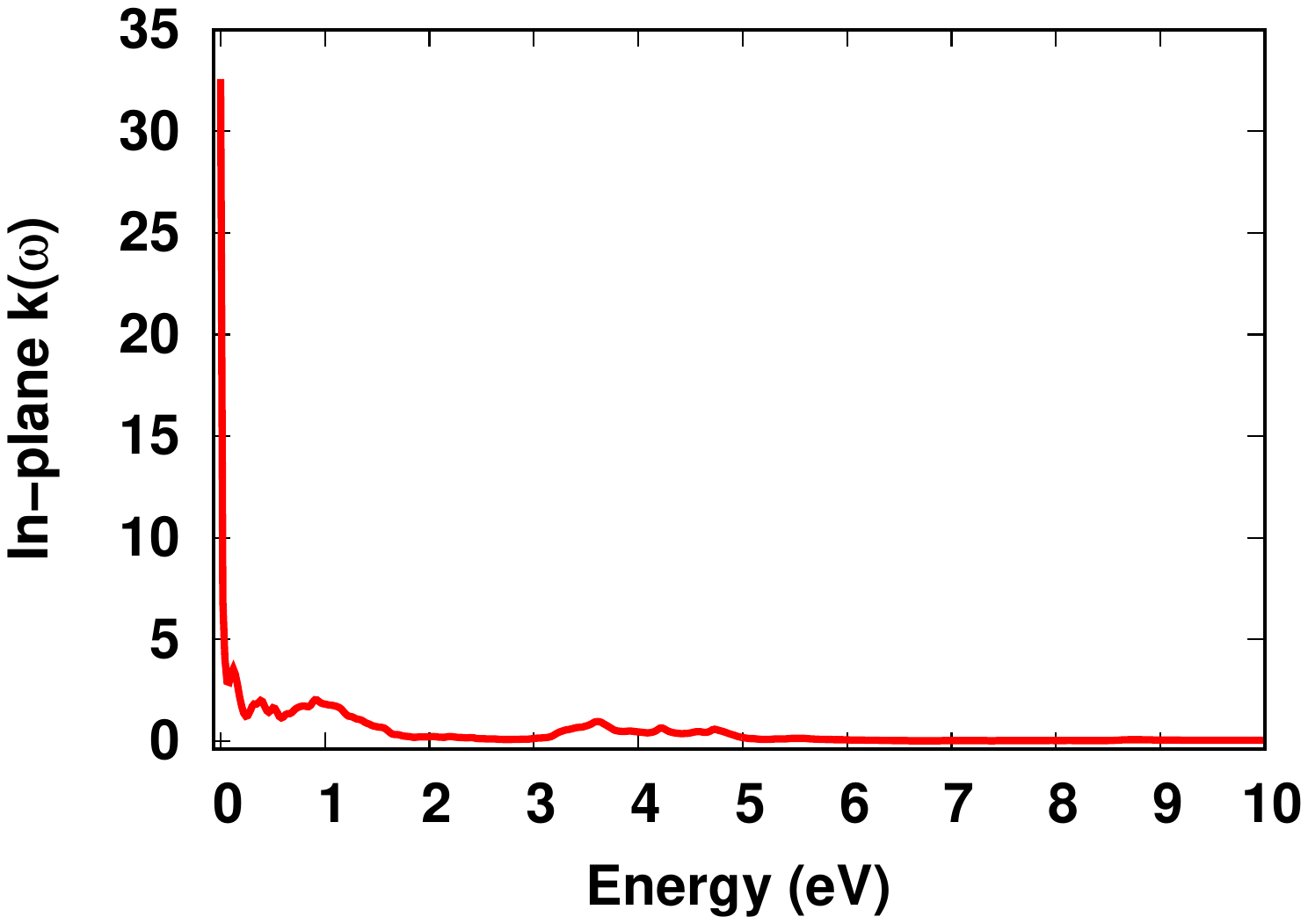}}
					\subfigure[]{\label{subfig:8(b)}
						\includegraphics[scale=0.31]{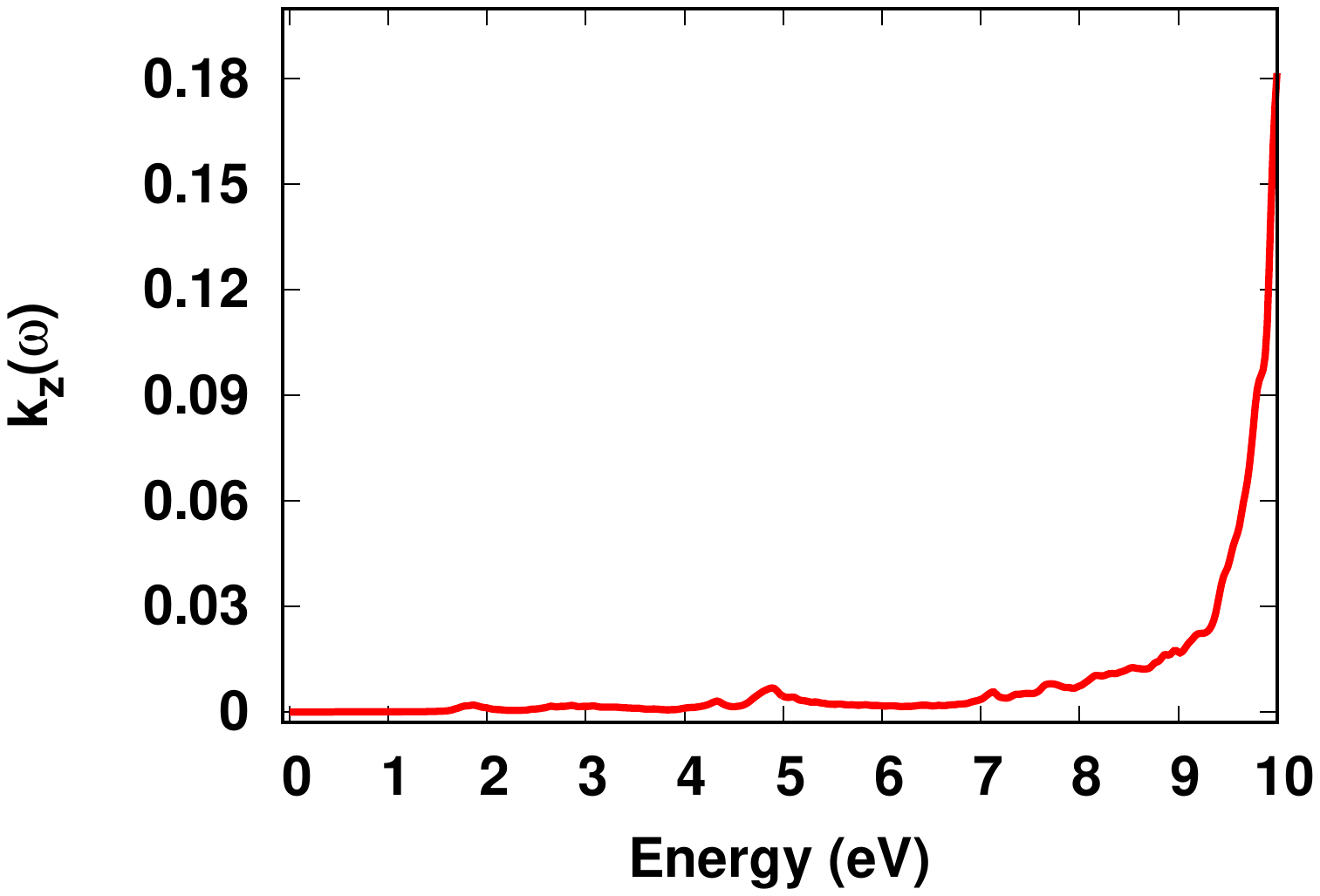}}
					\caption{\label{fig:8}
						Calculated extinction coefficient, $k(\omega)$, as a function of photon energy for $\alpha$-GDY. The (a) in-plane ($x-y$ plane) and (b) out-of-plane ($z$ direction, perpendicular to the monolayer plane) components are presented to illustrate the optical anisotropy.}
				\end{figure}
			\end{minipage}
		\end{widetext}
		The substantial difference between the in-plane and out-of-plane extinction coefficients further confirms the pronounced optical anisotropy of $\alpha$-GDY. The strong in-plane response is consistent with the highly delocalized electronic states within the carbon framework, whereas the much weaker out-of-plane response reflects the strongly anisotropic electronic polarizability of the monolayer. These results are consistent with the calculated dielectric function, absorption coefficient, and reflectivity spectra, providing a comprehensive description of the anisotropic optical response of $\alpha$-GDY.
		\subsubsection{Refractive index}
		Figure~\ref{fig:9} illustrates the calculated refractive index, $n(\omega)$, of $\alpha$-GDY for the in-plane and out-of-plane polarization directions.

		The refractive index exhibits a pronounced anisotropy over the entire investigated photon-energy range. The in-plane component is characterized by a very large value in the low-energy region, followed by a rapid decrease with increasing photon energy. Such behavior is a direct consequence of the metallic nature of $\alpha$-GDY and originates from the strong interaction of free charge carriers with low-energy electromagnetic radiation. As the photon energy increases, the refractive index gradually approaches a nearly constant value close to unity, indicating a significant reduction in the optical dispersion at higher photon energies.
		
		In contrast, the out-of-plane refractive index remains nearly constant over a wide energy range, with a value close to 1.05 at low photon energies. Above approximately $6$~eV, $n_z(\omega)$ increases gradually and reaches a maximum value of about 1.24 near $10$~eV. Compared with the in-plane response, the out-of-plane refractive index exhibits much weaker energy dependence, reflecting the limited electronic polarization perpendicular to the monolayer.
		
		The distinct difference between the two polarization directions demonstrates the intrinsic optical anisotropy of $\alpha$-GDY. The large in-plane refractive index at low photon energies is closely related to the free-carrier response and the strong dielectric screening within the carbon network, whereas the comparatively weak out-of-plane response arises from the reduced electronic polarizability normal to the atomic plane. These results are consistent with the calculated dielectric function, extinction coefficient, absorption coefficient, and reflectivity spectra, confirming the strongly anisotropic optical behavior of the $\alpha$-GDY monolayer.
						\begin{widetext}
			\begin{minipage}{\linewidth}
				\begin{figure}[H]
					\centering
					\subfigure[]{\label{subfig:9(a)}
						\includegraphics[scale=0.31]{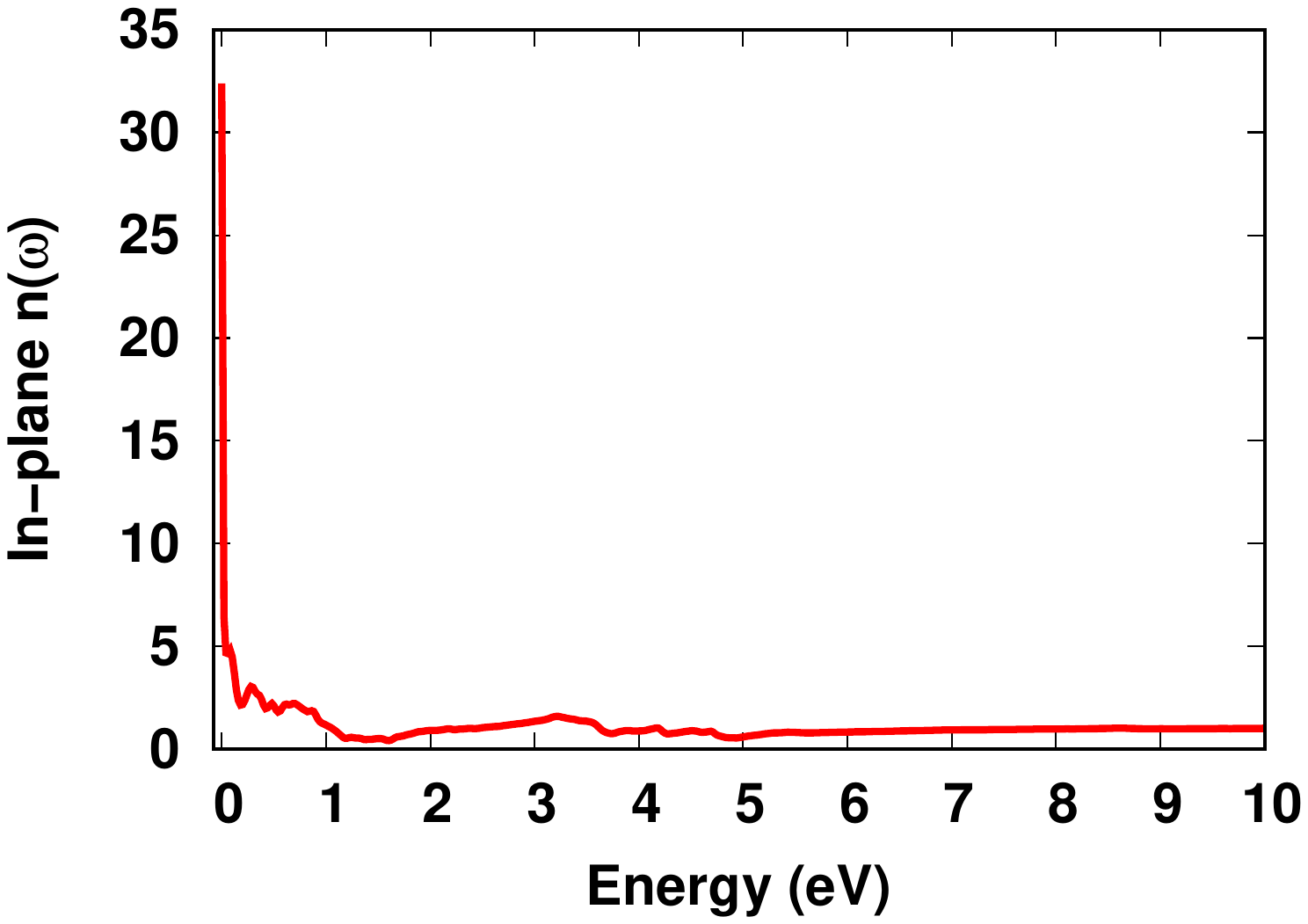}}
					\subfigure[]{\label{subfig:9(b)}
						\includegraphics[scale=0.31]{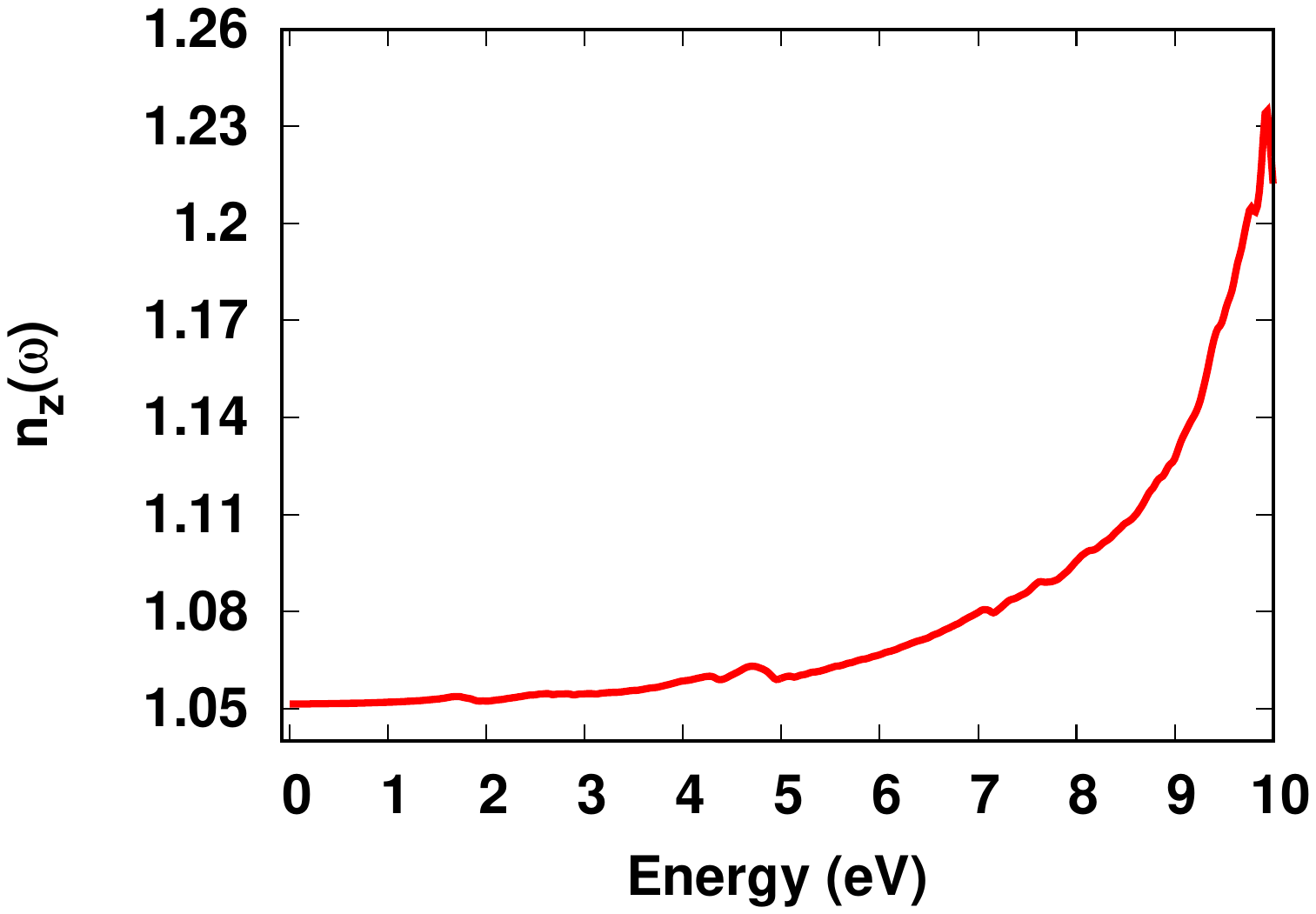}}
					\caption{\label{fig:9}
						Calculated refractive index, $n(\omega)$, as a function of photon energy for $\alpha$-GDY. The (a) in-plane ($x-y$ plane) and (b) out-of-plane ($z$ direction, perpendicular to the monolayer plane) components are shown for comparison.}
				\end{figure}
			\end{minipage}
		\end{widetext}
		\subsubsection{Electron energy-loss spectra (EELS)}
		Figure~\ref{fig:10} illustrates the calculated EELS, $\mathrm{Im}\!\left[-1/\varepsilon(\omega)\right]$, of $\alpha$-GDY for both in-plane and out-of-plane polarization directions. A pronounced optical anisotropy is observed throughout the investigated photon-energy range. The in-plane EELS spectrum exhibits several well-defined loss features, with the most prominent peaks located around $1.6$ and $4.9$~eV. Additional weaker features are observed between approximately $3$ and $5$~eV. These structures reflect enhanced electronic energy-loss channels arising from the frequency-dependent dielectric response of the monolayer.

		The calculated plasma frequencies obtained from the dielectric response are approximately $3.208$~eV along both the $x$ and $y$ directions, whereas a substantially smaller value of $1.059$~eV is obtained along the out-of-plane ($z$) direction. It should be noted, however, that the plasma frequency obtained from the dielectric response does not by itself provide an unambiguous assignment of individual peaks in the EELS spectrum. In particular, a rigorous identification of a loss feature as a collective plasmon requires consideration of the simultaneous behavior of $\varepsilon_1(\omega)$ and $\varepsilon_2(\omega)$, whereas interband transitions can also produce pronounced structures in the loss function. Accordingly, the prominent features around $1.6$ and $4.9$~eV are interpreted here as enhanced energy-loss features rather than being assigned exclusively to collective plasmons. Their spectral positions may reflect contributions from interband electronic transitions together with collective electronic excitations. The broad distribution of the loss features therefore reflects the combined influence of the electronic transitions and the dielectric response of the monolayer.
				\begin{widetext}
	\begin{minipage}{\linewidth}
		\begin{figure}[H]
			\centering
			\subfigure[]{\label{subfig:10(a)}
				\includegraphics[scale=0.31]{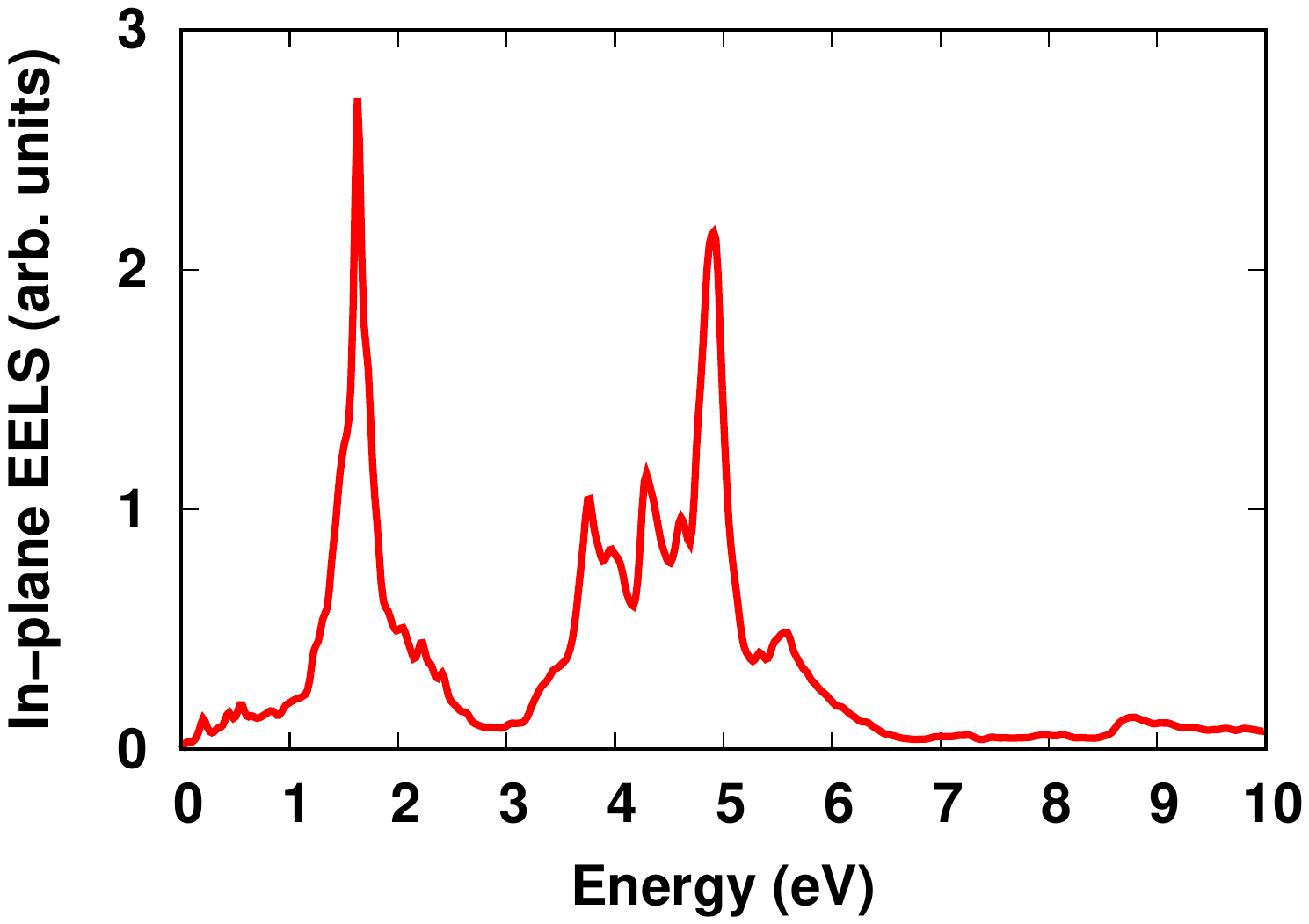}}
			\subfigure[]{\label{subfig:10(b)}
				\includegraphics[scale=0.31]{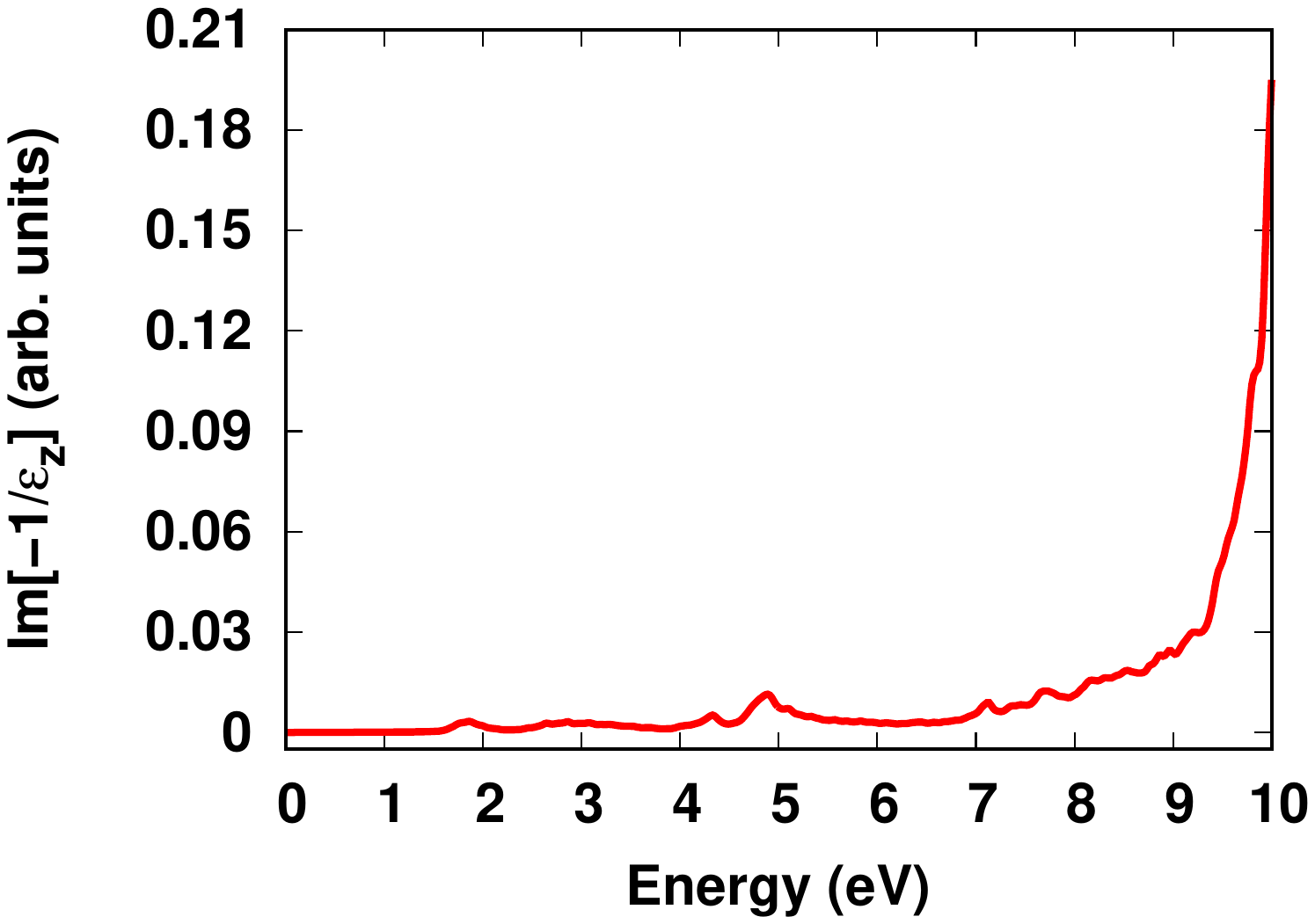}}
			\caption{\label{fig:10}
				Calculated electron energy loss spectra (EELS), $\text{Im}[-1/\epsilon(\omega)]$, as a function of photon energy for $\alpha$-GDY. The (a) in-plane ($x-y$ plane) and (b) out-of-plane ($z$ direction, perpendicular to the monolayer plane) components are presented to illustrate the optical anisotropy.}
		\end{figure}
	\end{minipage}
\end{widetext}		
		In contrast, the out-of-plane EELS spectrum, $\text{Im}[-1/\epsilon(\omega)]$, remains nearly zero over most of the investigated energy range and increases gradually only above approximately $8$~eV, reaching its maximum value near $10$~eV. Even at its maximum, the out-of-plane loss intensity is more than one order of magnitude smaller than that of the in-plane component, demonstrating the strongly suppressed out-of-plane electronic response.
		
		The pronounced difference between the in-plane and out-of-plane EELS spectra confirms the strong dielectric anisotropy of $\alpha$-GDY. The enhanced in-plane response is consistent with the delocalized electronic states within the carbon network, whereas the much weaker out-of-plane response reflects the strongly anisotropic electronic polarizability of the monolayer.
		\section{Conclusions}
		In this work, the structural, electronic, and optical properties of monolayer $\alpha$-graphdiyne ($\alpha$-GDY) were systematically investigated within the framework of density-functional theory. The optimized atomic structure preserves the characteristic hexagonal geometry of the monolayer. The calculated electronic band structure reveals a gapless Dirac crossing at the K point, indicating Dirac-semimetallic behavior within the PBE/GGA framework. The total and orbital-projected densities of states further characterize the electronic structure, showing that the states near the Fermi level are predominantly derived from carbon $2p$ orbitals, whereas the contribution of the $2s$ orbitals is comparatively weak.
		
		The calculated optical properties reveal pronounced anisotropy between the in-plane and out-of-plane polarization directions. The in-plane dielectric response exhibits a strong low-energy electronic response and negative values of its real part, whereas the out-of-plane component remains positive throughout the investigated energy range, with a static value of approximately $1.1$ and a gradual increase to about $1.5$ near $10$~eV. Similar anisotropic behavior is observed in the refractive index, extinction coefficient, absorption coefficient, reflectivity, and electron energy-loss spectra. The calculated plasma frequencies are approximately $3.21$~eV for the in-plane direction and $1.06$~eV for the out-of-plane direction, highlighting the strongly anisotropic electronic response of the monolayer. The prominent features observed in the EELS spectrum are interpreted as enhanced energy-loss channels that may contain contributions from both interband transitions and collective electronic excitations.
		
		It should be noted that the electronic and optical properties reported here are obtained within the PBE/GGA and independent-particle frameworks. Consequently, quantitative corrections to the band structure and optical excitation energies, as well as excitonic effects in the optical response, are not explicitly included. Higher-level approaches such as HSE06 or GW--BSE would be valuable for further validation of the electronic structure and optical spectra. Overall, the combination of Dirac-like electronic behavior and pronounced optical anisotropy highlights the potential relevance of $\alpha$-GDY for polarization-sensitive optoelectronic, plasmonic, and nanoelectronic applications.
		\section*{Data availability}
		The data supporting the findings of the present study are available within the article. Additional data are available from the corresponding author upon request.
		
		\end{document}